\documentclass[preprint,showpacs,preprintnumbers,amsmath,amssymb]{revtex4}

\usepackage{graphicx}% Include figure files
\usepackage{dcolumn}% Align table columns on decimal point
\usepackage{bm}% bold math

\begin{document}

\preprint{APS/123-QED}

\title{Lyapunov Analysis of Homogeneous Isotropic Turbulence}% Force line breaks with \\

\author{Nicola de Divitiis}
 \altaffiliation[ ]{via Eudossiana, 18,  00184, Rome}%Lines break automatically or can be 
 \email{dedivitiis@dma.dma.uniroma1.it}
\affiliation{%
Department of Mechanics and Aeronautics\\
University "La Sapienza", Rome, Italy  %\textbackslash\textbackslash
}%

\date{\today}

\begin{abstract}
The present work studies the isotropic and homogeneous turbulence for incompressible fluids through a specific Lyapunov analysis, assuming that the turbulence is due to the bifurcations associated to the velocity field.

The analysis consists in the study of the mechanism of the energy cascade from large to small scales through the Lyapunov analysis of the relative motion between two particles and in the
calculation of the velocity fluctuation through the Lyapunov analysis of the local deformation
and the Navier-Stokes equations.

The analysis provides an explanation for the mechanism of the energy cascade,
leads to the closure of the von K\'arm\'an-Howarth equation, and describes the 
statistics of the velocity difference.

Several tests and numerical results are presented.

\end{abstract}

\pacs{Valid PACS appear here}% PACS, the Physics and Astronomy
\maketitle

\newcommand{\no}{\noindent}
\newcommand{\be}{\begin{equation}}
\newcommand{\ee}{\end{equation}}
\newcommand{\bea}{\begin{eqnarray}}
\newcommand{\eea}{\end{eqnarray}}
\newcommand{\bc}{\begin{center}}
\newcommand{\ec}{\end{center}}

\newcommand{\calr}{{\cal R}}
\newcommand{\calv}{{\cal V}}

\newcommand{\bff}{\mbox{\boldmath $f$}}
\newcommand{\bfg}{\mbox{\boldmath $g$}}
\newcommand{\bfh}{\mbox{\boldmath $h$}}
\newcommand{\bfi}{\mbox{\boldmath $i$}}
\newcommand{\bfm}{\mbox{\boldmath $m$}}
\newcommand{\bfp}{\mbox{\boldmath $p$}}
\newcommand{\bfr}{\mbox{\boldmath $r$}}
\newcommand{\bfu}{\mbox{\boldmath $u$}}
\newcommand{\bfv}{\mbox{\boldmath $v$}}
\newcommand{\bfx}{\mbox{\boldmath $x$}}
\newcommand{\bfy}{\mbox{\boldmath $y$}}
\newcommand{\bfw}{\mbox{\boldmath $w$}}
\newcommand{\bfk}{\mbox{\boldmath $\kappa$}}

\newcommand{\bfA}{\mbox{\boldmath $A$}}
\newcommand{\bfD}{\mbox{\boldmath $D$}}
\newcommand{\bfI}{\mbox{\boldmath $I$}}
\newcommand{\bfL}{\mbox{\boldmath $L$}}
\newcommand{\bfM}{\mbox{\boldmath $M$}}
\newcommand{\bfS}{\mbox{\boldmath $S$}}
\newcommand{\bfT}{\mbox{\boldmath $T$}}
\newcommand{\bfU}{\mbox{\boldmath $U$}}
\newcommand{\bfX}{\mbox{\boldmath $X$}}
\newcommand{\bfY}{\mbox{\boldmath $Y$}}
\newcommand{\bfK}{\mbox{\boldmath $K$}}

\newcommand{\bfrho}{\mbox{\boldmath $\rho$}}
\newcommand{\bfchi}{\mbox{\boldmath $\chi$}}
\newcommand{\bfphi}{\mbox{\boldmath $\phi$}}
\newcommand{\bfPhi}{\mbox{\boldmath $\Phi$}}
\newcommand{\bflambda}{\mbox{\boldmath $\lambda$}}
\newcommand{\bfxi}{\mbox{\boldmath $\xi$}}
\newcommand{\bfLambda}{\mbox{\boldmath $\Lambda$}}
\newcommand{\bfPsi}{\mbox{\boldmath $\Psi$}}
\newcommand{\bfomega}{\mbox{\boldmath $\omega$}}
\newcommand{\bfeps}{\mbox{\boldmath $\varepsilon$}}
\newcommand{\bfkappa}{\mbox{\boldmath $\kappa$}}
\newcommand{\itPsi}{\mbox{\it $\Psi$}}
\newcommand{\itPhi}{\mbox{\it $\Phi$}}
\newcommand{\bint}{\mbox{ \int{a}{b}} }
\newcommand{\ds}{\displaystyle}
\newcommand{\Sum}{\Large \sum}

\section{\bf Introduction \label{s1} }

This work presents a study of isotropic and homogeneous turbulence 
for an incompressible fluid in an infinite domain.
The analysis is mainly  motivated by the fact that in turbulence the kinematics 
of the fluid deformation is subjected to bifurcations \cite{Landau44} and exhibits a chaotic 
behavior and huge mixing \cite{Ottino90}, resulting to be much more rapid than the fluid 
state variables.
This characteristics implies that the accepted kinematical hypothesis for deriving the Navier-Stokes equations could require the consideration of very small length scales and times for describing the fluid motion \cite{Truesdell77} and therefore a very large 
number of degrees of freedom. 
\\
Other peculiar characteristics of the turbulence are the mechanism of the kinetic energy cascade, directly related to the relative motion of a pair of fluid particles \cite{Richardson26, Kolmogorov41, Karman38, Batchelor53} and responsible for the shape of 
the developed energy spectrum, and the non-gaussian statistics of the velocity difference.

The present analysis assumes that the fluctuations of all the fluid state variables
are the result of the bifurcations of the velocity field. The evolution in the time of these fluctuations is calculated with the Lyapunov analysis of the particle equations of motion.

The first part of the work deals with the representation of velocity difference
between two fixed points of the space. This is analyzed with the Lyapunov theory
 studying the motion of the particles crossing the two points. 
This analysis gives an explanation of the mechanism of kinetic energy transfer between length scales and leads to the closure of the von K\'arm\'an-Howarth equation \cite{Karman38}. The obtained expression of the function $K(r)$, which represents the inertia forces, is in terms of the longitudinal correlation function and its spatial derivative, and  satisfies the conservation law which states that the inertia forces only transfer the kinetic energy \cite{Karman38, Batchelor53}.

In the second part, the statistics of the velocity difference is studied through the kinematics
of the local deformation and the momentum equations. 
These momentum equations are expressed with respect to the referential coordinates which coincide with the material coordinates for a given fluid configuration \cite{Truesdell77}, whereas the kinematics of the local deformation is analyzed with the Lyapunov theory.
The choice of the referential coordinates allows the velocity fluctuations to be analytically expressed in terms of the Lyapunov exponent of the local fluid deformation.
The statistics of velocity difference is studied with the Fourier analysis of the velocity fluctuations, and an analytical expression for the velocity difference and for its PDF
is obtained in case of isotropic turbulence. This expression incorporates an unknown function,
related to the skewness, which is identified through the obtained expression of $K(r)$.
The velocity difference also requires the knowledge of the critical Reynolds number whose estimation is made in the Appendix B, where the order of magnitude is roughly determined through a qualitative analysis of the bifurcations of the velocity field. 

Finally, the several results obtained with this analysis are compared with the data
existing in the literature, indicating that the proposed analysis can adequately describe the
various properties of the fully developed turbulence.

The present analysis only studies the possibility to obtain the fully developed homogeneous-isotropic turbulence in a given condition and does not analyze the intermediate stages of the turbulence.

\section{\bf Lyapunov analysis of the relative motion:
 Closure of the von K\'arm\'an-Howarth equation  \label{s5}}

\begin{figure}
\vspace{-0.mm}
\centering
 \includegraphics[width=0.4\textwidth]{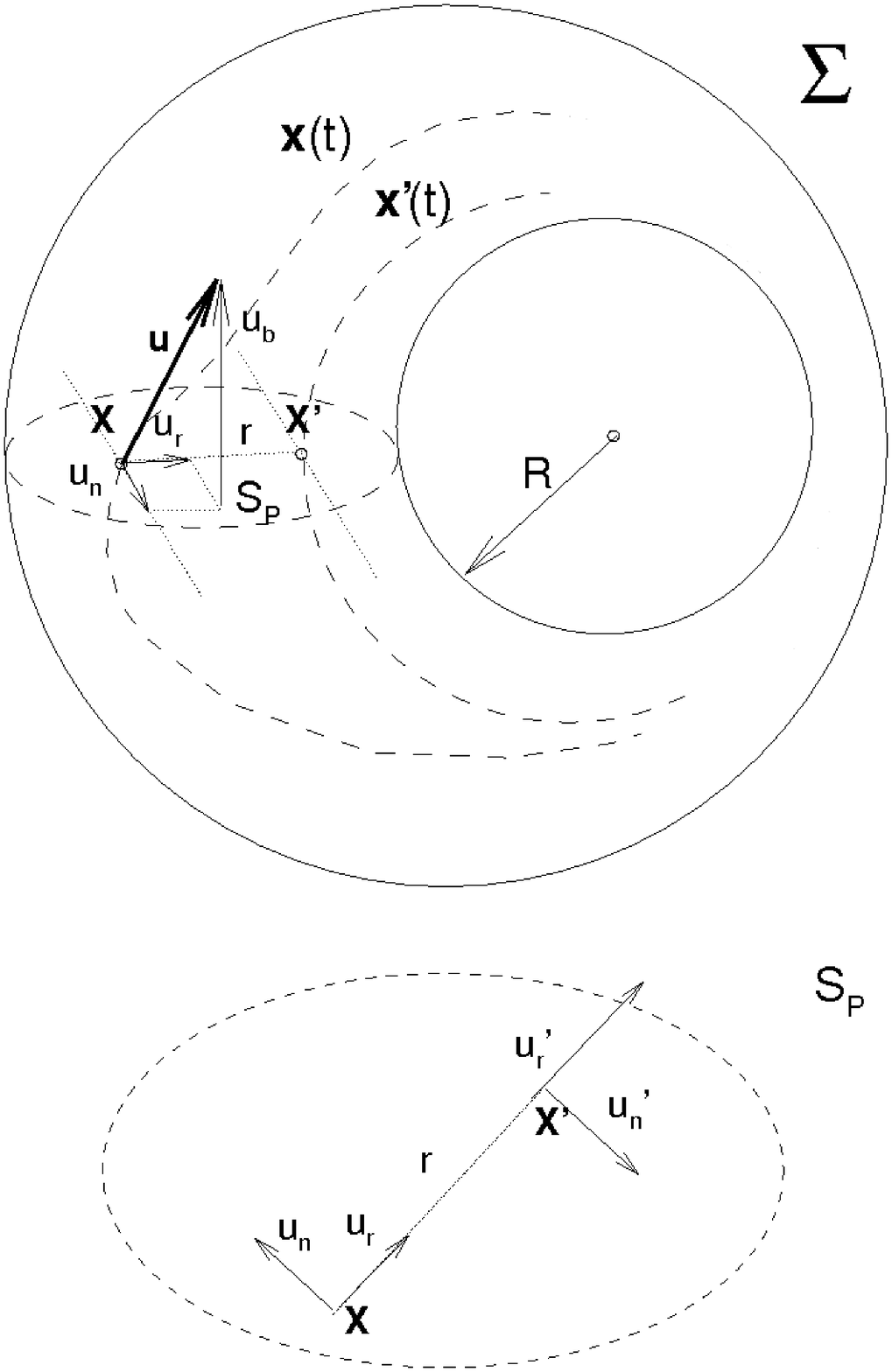}
\vspace{-0.mm}
\caption{Scheme of the relative motion of two fluid particles}
\vspace{-0.mm}
\label{figura_1a}
\end{figure}

In order to investigate the mechanism of the energy cascade, the properties of the relative equations of motion between fluid particles are here studied with the Lyapunov analysis.
To this purpose, consider two fixed points of the space, ${\bf X}$ and ${\bf X}'$ (see Fig. \ref{figura_1a}) whose distance is $r = \vert {\bf X}'-{\bf X} \vert$ and the motion 
of two fluid particles which at a given time $t_0$, cross through ${\bf X}$ and ${\bf X}'$. 
The equations of motion of these particles are
\bea
\begin{array}{l@{\hspace{+0.2cm}}l}
\ds \frac{d {\bf x}}{d t} = {\bf u} ({\bf x}, {t}),  \ \  \frac{d {\bf x}'}{d t}
 = {\bf u} ({\bf x}', {t})
\end{array}
\label{k_1}
\label{k_0}
\eea 
At $t_0$, a toroidal volume $\Sigma (t_0)$ is chosen which contains ${\bf X}$ and ${\bf X}'$,
whose geometry and position change according to the fluid motion. 
In Fig. \ref{figura_1a},  $S_p \gtrapprox r^2$ and $R$ are, respectively, the poloidal surface and the toroidal dimension of $\Sigma$ which vary with time to preserve the volume.
The velocity difference associated to ${\bf X}$ and ${\bf X}'$ is 
$\Delta{\bf u} ={\bf u}({\bf X'}, t)-{\bf u}({\bf X}, t)$ 
and its components $\Delta u_n \equiv u_n'-u_n$ and 
$\Delta u_r \equiv u_r'-u_r$, lay on $S_p$ and are normal and parallel to $r$, respectively, whereas $u_b$ is the average of the velocity components along the direction normal to $S_p$.
According to the theory \cite{Guckenheimer90}, for $t > t_0$, the trajectories of the two particles are enclosed into $\Sigma (t)$. For sake of simplicity and without loss of generality, we assume that $R$ increases with time \cite{Guckenheimer90}. 
The Lyapunov analysis of Eqs. (\ref{k_0}) leads to  
\bea
\ds R \approx  R(t_0) \ \mbox{e}^{\lambda (t-t_0)}
\label{Lyapunov}
\eea
These variations of $R$ are caused of the bifurcations \cite{Guckenheimer90} 
of Eqs. (\ref{k_0}).
Since $R$ rises with time, $\lambda(r) >$ 0 identifies the maximal finite scale Lyapunov exponent associated to Eqs. (\ref{k_0}).

The equations of motion for $\ds \Sigma(t)$ preserve the volume and can be expressed in terms of the velocity components calculated at $\bf X$ and $\bf X'$ \cite{Lamb45}. These are
\bea
\begin{array}{l@{\hspace{+0.0cm}}l}
\ds \frac{\partial}{\partial t} \left( S_p R  \right) = 0 
\end{array}
\label{laws0}
\eea
\bea
\begin{array}{l@{\hspace{+0.0cm}}l}
\ds \frac{\partial}{\partial t} \left( (\Delta u_n)^2 S_p \right) = J S_p \\\\
\ds \frac{\partial}{\partial t} \left( u_b R  \right) = -\nu \beta \frac{u_b}{R}
\end{array}
\label{laws1}
\eea
In line with Lamb \cite{Lamb45}, Eqs (\ref{laws0}) and (\ref{laws1}) represent, respectively, the continuity equation and the momentum equations   which can be derived from the integral equations of balance over $\Sigma$. 
Into Eq.(\ref{laws1}), $\nu$ is the kinematic viscosity, $\beta = O(1) >0$ is a proper
constant, and $J$ is related to the time derivative of the kinetic energy and to the viscosity \cite{Lamb45}. $J$ is equal to zero when $\nu =$ 0.

Equations (\ref{laws0}) and (\ref{laws1}) are written in terms of the fluid properties 
calculated at $\bf X$ and $\bf X'$, thus  are referred to the Eulerian description of motion \cite{Truesdell77, Lamb45}.
Substituting  Eq. (\ref{Lyapunov}) into 
%the result from the Lyapunov analysis of Eq. (\ref{k_0}), 
Eqs. (\ref{laws0}) and (\ref{laws1}), one obtains
\bea
\begin{array}{l@{\hspace{+0.0cm}}l}
\ds \frac{\partial  (\Delta u_r)^2}{\partial t} = - \lambda (\Delta u_r)^2 \\\\
\ds \frac{\partial (\Delta u_n)^2}{\partial t} =  \lambda (\Delta u_n)^2 + J \\\\
\ds \frac{\partial u_b}{\partial t} =  - (\lambda   + \beta \frac{\nu}{R^2}) u_b
\end{array}
\label{laws-lyap}
\eea
Since $\lambda >0$, $u_b \rightarrow 0$.
Equations (\ref{laws-lyap}) describe fluctuations of velocity difference caused by bifurcation of Eqs. (\ref{k_0}) 
and hold as long as $\bf X$ and $\bf X'$ are both enclosed into $\Sigma (t)$.
This condition is verified if $t - t_0$ does not exceed very much the order of magnitude of $1/\lambda$ \cite{Guckenheimer90}.

\bigskip

In order to obtain the closure of the von K\'arm\'an-Howarth equation,
Eq. (\ref{laws-lyap}) enter the computation of  the average of the physical
quantity $\Upsilon$:
\bea
\ds \frac{\partial}{\partial r_i}  \left( \Upsilon  r_i \right) \equiv 
%\Sum_{i = r, n, b} 
\frac{\partial }{\partial t} \left( u_i u'_i \right) 
- \frac{1}{\rho}\frac{\partial T_{i j}}{\partial x_j}u'_i
- \frac{1}{\rho}\frac{\partial T'_{i j}}{\partial x'_j}u_i
\label{Upsilon}
\eea
where $T_{i j}$  =  $ \ds - {p} \delta_{i j} + 
\nu \rho \left(  {\partial u_i}/{\partial x_j} + {\partial u_j}/{\partial x_i} \right) 
$ is the stress tensor. The repeated indexes  denote the summation
with respect to the same indexes, which are $i$ = $r, n, b$ and $j$ = $r, n, b$.

According to von K\'arm\'an \cite{Karman38}, $\Upsilon$ expresses that part of the
inertia forces, responsible for the transferring of the kinetic energy between
the several fluid regions, whose average only depends on the current value of the average
kinetic energy.
In the von K\'arm\'an-Howarth equation, the function $K(r)$ is the average of $\Upsilon$.
The average is calculated on all the pairs of particles which cross through ${\bf X}$ and ${\bf X}'$ at the same time. Specifically, $K(r)$ is determined substituting Eqs. (\ref{laws-lyap}) into Eq. (\ref{Upsilon}), assuming the homogeneity and the isotropy and
taking into account that  $\langle \Upsilon \rangle$ does not depend neither on  
$\partial \langle u_i u_i \rangle / \partial t$, nor  on $\langle {\partial T_{i j}}/{\partial x_j}u'_i
+ {\partial T'_{i j}}/{\partial x'_j}u_i \rangle$ \cite{Karman38, Batchelor53}.
This immediately identifies
\bea
\begin{array}{l@{\hspace{+0.0cm}}l}
K(r) \equiv
\langle \Upsilon \rangle   =
\ds \lambda \ u^2 \ ( g - f ) \\\\
\ds \langle J \rangle = 4 \ \frac{\partial u^2}{ \partial t} - \frac{2}{\rho}
 \langle \frac{\partial T_{i j}}{\partial x_j}u'_i + \frac{\partial T'_{i j}}
{\partial x'_j}u_i \rangle
\end{array}
\label{vv'}
\eea
where $u^2=\langle u_i u_i\rangle /3$ and $f$ and $g$ are longitudinal and lateral velocity correlation functions. Due to the fluid incompressibility, $f$ and $g$ are related each other through $g = f +1/2 \partial f /\partial r$ (see Eq. (\ref{g}), Appendix A), leading
to the expression
\bea
K(r) =
\frac{1}{2} u^2    \frac{\partial f}{\partial r} \ \lambda(r) r
\label{vv'1}
\eea

Considering that $K(r)$ does not directly depend on the viscosity, this expression can be also obtained at $\nu$ = 0. 
In this case $\langle J  \rangle$ = 0, $\partial \langle u_i u_i \rangle /\partial t$ = 0,  
$\langle {\partial T_{i j}}/{\partial x_j}u'_i + {\partial T'_{i j}}/{\partial x'_j}u_i \rangle$ = 0 \cite{Karman38}, and Eqs. (\ref{vv'}) and (\ref{vv'1}) are again recovered.

\bigskip

Equation (\ref{vv'1}) states that, the fluid incompressibility, expressed by 
$g - f \ne 0$, 
represents a sufficient condition to state that $K(r) \ne$ 0. This latter
is determined as soon as $\lambda$ is known. 
To calculate $\lambda$, it is convenient to express 
$\Delta {\bf u} = {\bf u}({\bf X}', t) - {\bf u}({\bf X}, t)$ in the Lyapunov basis 
of orthonormal vectors $E \equiv({\bfeps}_1, {\bfeps}_2, {\bfeps}_3)$
associated to Eqs. (\ref{k_1}) \cite{Christiansen97, Ershov98}.
The velocity difference expressed in $E$, 
$\Delta {\bf v} \equiv (v_1' -v_1, \ v_2' -v_2, \ v_3' -v_3 )$,  
satisfies the following equations, which hold for times whose order of magnitude do not exceed very much $1/{\lambda}$ \cite{Guckenheimer90, Prigogine94}
\bea 
\begin{array}{l@{\hspace{+0.0cm}}l}
\ds v_i' -v_i   =  {\lambda}_ i \ \hat{r}_i, \ \ \ i = 1, 2, 3 
\end{array}
\label{Lyap}
\eea
where $\hat{r}_i$,  $v_i$ and  $v_i'$ are, respectively, the components of
 ${\bf X}'-{\bf X}$, ${\bf u} ({\bf X}, t) $ and ${\bf u} ({\bf X}', t)$ written in $E$.
Then, $r$ and $\Delta u_r$  can be expressed in terms of $\hat{\bf r}$ and $\Delta {\bf v}$ as
\bea
\begin{array}{l@{\hspace{+0.0cm}}l}
r =  {\bfxi} \cdot {\bf Q} \hat{\bf r},  \ \ \
\Delta u_r = {\bfxi} \cdot   {\bf Q} \Delta {\bf v}
\end{array}
\label{rot_E_R}
\eea
Into Eqs. (\ref{rot_E_R}), ${\bf Q} \equiv (( \varepsilon_{i j}))$ is the rotation matrix transformation from $E$ to $\Re$, where $\varepsilon_{i j}$ is the component of ${\bfeps}_j$ along the coordinate direction $i$ on $\Re$, and 
${\bfxi} = ({\bf X}'-{\bf X})/\vert {\bf X}'-{\bf X} \vert $.

The standard deviation of $\Delta u_r$ is calculated from Eqs. (\ref{rot_E_R}), taking into 
account that $\Delta {\bf v} \approx \lambda  \hat{\bf r}$ and that $\bf Q$
is fluctuating depending on the pair paths:
\bea
\left\langle (\Delta u_r)^2 \right\rangle = 
\sum_{i=1}^3 \sum_{j=1}^3 \sum_{p=1}^3 \sum_{q=1}^3 \xi_i \xi_p
\langle {\lambda}^2  \varepsilon_{i j} \varepsilon_{p q} \rangle r_j r_q
%\nonumber
\label{1A_1}
\eea 
Since $\lambda$ is calculated as the average of the velocity increment per unit distance, it is constant with respect the statistics of  $\varepsilon_{i j}$ and  $\varepsilon_{p q}$ \cite{Ershov98}, thus
$\langle {\lambda}^2  \varepsilon_{i j} \varepsilon_{p q} \rangle =
{\lambda}^2 \langle  \varepsilon_{i j} \varepsilon_{p q} \rangle$.
Furthermore, due to isotropy, the Lyapunov vectors fluctuate in such a way that 
$\langle  \varepsilon_{i j} \varepsilon_{p q} \rangle = \delta_{i j} \delta_{p q}$\cite{Ershov98}.
As the result, the standard deviation of the longitudinal velocity difference is
\bea
\left\langle (\Delta u_r)^2 \right\rangle = {\lambda}^2 r^2
\label{1A}
\eea 
This standard deviation can be also expressed through the longitudinal correlation function $f$
\bea
\langle (\Delta u_r)^2 \rangle = 2 u^2 (1-f(r))
\label{1B}
\eea
being $u$ the standard deviation of the longitudinal velocity.
The maximal Lyapunov exponent is calculated in function of $f$,
from Eqs. (\ref{1A}) and (\ref{1B})
\bea
\ds {\lambda} (r) = \frac{u}{r} \sqrt{2 \left( 1-f(r) \right) }
\label{lC}
\eea
Hence, substituting Eq. (\ref{lC}) into Eq. (\ref{vv'1}), one obtains
the expression of $K(r)$ in terms of the longitudinal correlation function
\bea
K(r) =
u^3 \sqrt{\frac{1-f}{2}} \ \frac{\partial f}{\partial r} 
\label{vv'2}
\label{vk6}
\eea
Thanks to the isotropy, $K(r)$ is a function of $r$ alone.

Equation (\ref{vk6}) represents the proposed closure of the von K\'arm\'an-Howarth equation.
This is the consequence of the fact that, 
the kinetic energy, initially enclosed into $\Sigma(t_0)$, at the end of the fluctuation is contained into $\Sigma(t)$ whose dimensions are changed with respect to $\Sigma(t_0)$ in agreement with the Lyapunov theory. This corresponds to a mechanism of the kinetic energy  transferring between  diverse regions of space which 
preserves the average values of the momentum and of the  kinetic energy.
Specifically, the analytical structure of Eq.(\ref{vk6}) states that this mechanism consists of a flow of the kinetic energy from large to small scales which only redistributes the kinetic energy between wavelengths.

\section{\bf skewness of velocity difference PDF \label{s6} }

The obtained expression of $K(r)$ allows to determine the skewness of $\Delta u_r$ \cite{Batchelor53}
\bea
\ds H_3(r) = \frac{\left\langle (\Delta u_r)^3 \right\rangle} 
{\left\langle (\Delta u_r)^2\right\rangle^{3/2}} =
  \frac{6 k(r)}{\left( 2 (1 -f(r)  )   \right)^{3/2} }
\label{H_3_01}
\eea
which is expressed in terms of the longitudinal triple correlation $k(r)$, linked to $K(r)$
by  $K(r)= u^3 \left( {\partial}/{\partial r}  + 4/r  \right) k(r)$ (also see Appendix A, 
Eq. (\ref{kk})).
Since $f$ and $k$ are, respectively, even and odd functions of $r$ with
 $f(0)$ = 1, $k(0) = k'(0)=k''(0)$ =0,   $ H_3(0)$ is given by 
 \bea
\ds H_3(0) = \lim_{r\rightarrow0} H_3(r) = \frac{k'''(0)}{(-f''(0))^{3/2}} 
\label{H_3_0}
\eea
where the apex denote the derivative with respect to $r$.
To obtain $H_3(0)$, observe that, near the origin, $K$ behaves as
\bea
\begin{array}{l@{\hspace{+0.cm}}l}
 \ds K = u^3 \sqrt{-f''(0)} f''(0) \frac{r^2}{2} + O(r^4)
\end{array}
\label{K0}
\eea
then, substituting Eq. (\ref{K0}) into $K(r)= u^3 \left( {\partial}/{\partial r}  + 4/r  \right) k(r)$ and accounting for Eq. (\ref{H_3_0}), one obtains
\bea
\ds H_3(0)  = -\frac{3}{7} = -0.42857...
\label{sk0}
\eea
This value of  $H_3(0)$ is a constant of the present analysis, 
which does not depend on the Reynolds number. 
This is in agreement with the several sources of data existing in the literature
such as \cite{Batchelor53, Tabeling96, Tabeling97, Antonia97} (and Refs. therein) and its  value gives the entity of the mechanism of energy cascade.

\section{\bf Statistical analysis of velocity difference \label{s4}}

As explained in this section, the Lyapunov analysis of the local deformation
and some plausible assumptions about the statistics of velocity difference
$\Delta {\bf u} ({\bf r}) \equiv {\bf u} ({\bf X}+{\bf r})-{\bf u} ({\bf X})$
lead to determine all the statistical moments of $\Delta {\bf u} ({\bf r})$
with only the knowledge of the function $K(r)$ and of the value of the critical
Reynolds number.

Starting from the momentum Navier-Stokes equations
\bea
\ds \frac{\partial  {u}_k}{\partial t}  = 
- \frac{\partial  {u}_k}{\partial x_h} u_h +
\frac{1}{\rho}  \frac{\partial T_{k h}}  {\partial x_{h}} 
\label{N-S}
\eea
consider the map $\bfchi$ : ${\bf x}_0 \rightarrow {\bf x}$, 
which is the function that determines the current position $\bf x$ of a fluid particle 
located at the referential position ${\bf x}_0$ \cite{Truesdell77} at $t = t_0$.
Equation (\ref{N-S}) can be written in terms of the referential position ${\bf x}_0$ \cite{Truesdell77} 
\bea
\ds \frac{\partial  {u}_k}{\partial t}  =  \left( -\frac{\partial  {u}_k}{\partial x_{0 p}} u_h +
\frac{1}{\rho}
 \frac{\partial T_{k h}}  {\partial x_{0 p}} \right)  \ \frac{\partial x_{0 p}}{\partial x_{h}} 
\label{N-Sr}
\eea
The Lyapunov analysis of the fluid strain provides the expression of this deformation in terms of the maximal Lyapunov exponent 
\bea
\frac{\partial {\bf x}}{\partial {\bf x}_0} \approx {\mbox e}^{\Lambda (t - t_0)} 
\label{stretch}
\eea
where $\Lambda \equiv \lambda(0) = \max (\Lambda_1, \Lambda_2, \Lambda_3 )$ is the maximal Lyapunov exponent and $\Lambda_i$, $(i = 1, 2, 3)$ are the Lyapunov exponents.
Due to the incompressibility, $\Lambda_1 + \Lambda_2 + \Lambda_3 =$ 0,
thus,  $\Lambda >0$.

If we assume that this deformation is much more rapid than 
$ {\partial T_{k h}} / {\partial x_{0 p}}$ and ${\partial  {u}_k}/{\partial x_{0 p}} u_h$,
the velocity fluctuation can be obtained from 
Eq. (\ref{N-Sr}), where ${\partial T_{k h}} / {\partial x_{0 p}}$ and
${\partial  {u}_k}/{\partial x_{0 p}} u_h$ are supposed to be constant
with respect to the time 
\bea
\begin{array}{l@{\hspace{0cm}}l}
\ds u_k \approx 
\frac{1}{\Lambda}  \ 
\left( -\frac{\partial  {u}_k}{\partial x_{0 p}} u_h +
\frac{1}{\rho}
 \frac{\partial T_{k h}}  {\partial x_{0 p}} \right)_{t = t_0} 
 \ds \approx  \frac{1}{\Lambda} \left(  \frac{\partial u_k} {\partial t} \right)_{t = t_0} 
\end{array}
\label{fluc_v2_0}
\label{fluc_v2}
\eea
This assumption is justified by the fact that, according to Truesdell \cite{Truesdell77},  ${\partial T_{k h}} / {\partial x_{0 p}} -{\partial  {u}_k}/{\partial x_{0 p}} u_h$ is a smooth function of time -at least during the period of a fluctuation-  whereas the fluid deformation varies very rapidly in proximity of a bifurcation according to Eq. (\ref{stretch}). 

\bigskip

The statistical properties of $\Delta {\bf u} ({\bf r})$, are  investigated expressing the velocity fluctuation, given by Eq. (\ref{fluc_v2}), as the Fourier series
\bea
\ds {\bf u} 
\approx
 \frac{1}{\Lambda} \sum_{\bfkappa} \frac{{\bf \partial U}}{\partial t} ({\bfkappa}) 
 {\mbox e}^{i {\bfkappa}\cdot {\bf x}}  
\label{f1}
\eea
where ${\bf U}({\bfkappa})$ $\equiv$ $(U_1({\bfkappa}), U_2({\bfkappa}), U_3({\bfkappa}))$ are the components of velocity spectrum, which satisfy the Fourier transformed Navier-Stokes equations \cite{Batchelor53}
\bea
\begin{array}{l@{\hspace{0.cm}}l}
\ds  \frac{\partial  U_p ({\bfkappa})}{\partial t}  = - \nu k^2 U_p ({\bfkappa}) + \\\\
\ds i \sum_{\bf j} ( \frac{\kappa_p \kappa_q \kappa_r}{\kappa^2}   U_q({\bf j}) 
U_r({\bfkappa} -{\bf j}) 
\ds - \kappa_q  U_q({\bf j}) U_p({\bfkappa} -{\bf j}) ) 
%\ d {\bf j}
\end{array}
\label{NS_fourier}
\eea
All the components ${\bf U}({\bfkappa}) \approx  \partial {\bf U}({\bfkappa})/ {\partial t} /\Lambda$ are random variables distributed according to certain distribution functions, which are statistically orthogonal each other \cite{Batchelor53}.

Thanks to the local isotropy, $\bf u$ is sum of several dependent random variables which are identically distributed \cite{Batchelor53}, therefore $\bf u$ tends to a gaussian variable \cite{Lehmann99}, and ${\bf U}({\bfkappa})$ satisfies the Lindeberg condition, a very general necessary and sufficient condition for satisfying the central limit theorem \cite{Lehmann99}. 
This condition does not apply to the Fourier coefficients of 
$\Delta {\bf u}$. In fact, since $\Delta {\bf u}$ is the difference between two dependent gaussian variables, its PDF could be a non gaussian
distribution function.
In ${\bf x}=0$, the velocity difference $\Delta {\bf u} ({\bf r}) \equiv
(\Delta u_1, \Delta u_2, \Delta u_3)$ is given by
\bea
\Delta u_p \hspace{-1.mm} \approx \hspace{-1.mm}   \frac{1}{\Lambda} 
\sum_{\bfkappa} \frac{\partial  U_p ({\bfkappa})} {\partial t} 
({\mbox e}^{i {\bfkappa}\cdot {\bf r}} - 1)    \equiv L + B + P + N
\eea
This fluctuation consists of the contributions appearing into Eq. (\ref{NS_fourier}):
in particular, $L$ represents the sum of all linear terms due to the viscosity
and $B$ is the sum of all bilinear terms arising from inertia and pressure
 forces. $P$ and $N$  are, respectively, the sums of definite positive 
and negative square terms, which derive from inertia and pressure forces.
The quantity $L+B$ tends to a gaussian random variable being 
the sum of statistically orthogonal terms \cite{Madow40, Lehmann99}, while $P$ and $N$ do
 not, as they are linear combinations of squares \cite{Madow40}.
 Their general  expressions are  \cite{Madow40}
\bea
\begin{array}{l@{\hspace{+0.2cm}}l}
 P = P_0 + \eta_1  +  \eta_2^2   \\\\
 N = N_0 + \zeta_1 +  \zeta_2^2  
\end{array} 
\label{nn}
\eea
where $P_0$ and $N_0$ are constants, and $\eta_1$, $\eta_2$, $\zeta_1$ and  $\zeta_2$ are four different centered random gaussian variables. 
Therefore, the fluctuation $\Delta u_r$ of the longitudinal velocity difference
can be written as
\bea
\begin{array}{l@{\hspace{+0.2cm}}l}
\ds \Delta {u}_p  = 
\psi_1({\bf r}) {\xi} + \psi_2({\bf r})  
\ds \left( \chi  ( {\eta}^2-1 )  -  ( {\zeta}^2-1 )  \right) 
\end{array}
\label{fluc3}
\eea
where $\xi$, ${\eta}$ and $\zeta$ are independent centered random variables
which have gaussian distribution functions with standard deviation equal to the unity.
The parameter $\chi$ is a positive definite function of Reynolds number, 
whereas $\psi_1$ and $\psi_2$ are functions of space
coordinates and the Reynolds number. 

At the Kolmogorov scale $\ell$, the order of magnitude of the velocity fluctuations is ${u_K}^2 \tau/\ell$, with $\tau = 1/\Lambda$ and $u_K = \nu / \ell$, whereas $\psi_2$ is negligible because is due to the inertia forces: this immediately identifies $\psi_1 \approx {u_K}^2 \tau/\ell$.
\\
On the contrary, at the Taylor scale $\lambda_T$, $\psi_1$ is negligible and the order of magnitude of the velocity fluctuations is $u^2 \tau/\lambda_T$, therefore $\psi_2 \approx u^2 \tau/\lambda_T$.

The ratio $\psi_2 / \psi_1$ is a function of $R_\lambda$
\bea
\psi({\bf r}, R_{\lambda}) = \frac{\psi_2 ({\bf r})}{\psi_1({\bf r})} 
\approx \frac{u^2 \ell}{{u_K}^2 \lambda_T} =  
\sqrt{\frac{R_{\lambda}}{15 \sqrt{15}}} \
\hat{\psi}({\bf r})
\label{Rl}
\eea
where 
$
\ds \hat{\psi}({\bf r}) = O(1)
\label{R2}
$, 
is a function which has to be determined.

Hence, the dimensionless longitudinal velocity difference $\Delta {u}_r$, is written as
\bea
\begin{array}{l@{\hspace{+0.2cm}}l}
\ds \frac {\Delta {u}_r}{\sqrt{\langle (\Delta {u}_r)^2} \rangle} =
\ds \frac{   {\xi} + \psi \left( \chi ( {\eta}^2-1 )  -  
\ds  ( {\zeta}^2-1 )  \right) }
{\sqrt{1+2  \psi^2 \left( 1+ \chi^2 \right)} } 
\end{array}
\label{fluc4}
\eea 
The dimensionless statistical moments of $\Delta {u}_r$
are easily calculated considering that $\xi$, $\eta$ and 
$\zeta$ are independent gaussian variables
\bea
\begin{array}{l@{\hspace{+0.2cm}}l}
\ds H_n \equiv \frac{\left\langle (\Delta u_r)^n \right\rangle}
{\left\langle (\Delta  u_r)^2\right\rangle^{n/2} }
= 
\ds \frac{1} {(1+2  \psi^2 \left( 1+ \chi^2 \right))^{n/2}} \\\\
\ds \sum_{k=0}^n 
 \binom{n}{k} \psi^k
 \langle \xi^{n-k} \rangle 
  \langle (\chi(\eta^2 -1) - (\zeta^2 -1 ) )^k \rangle 
\end{array}
\label{m1}
\eea
where
\bea
\begin{array}{l@{\hspace{+0.2cm}}l}
\ds   \langle (\chi(\eta^2 -1) - (\zeta^2 -1 ) )^k \rangle = \\\\
\ds \sum_{i=0}^k 
 \binom{k}{i} (-\chi)^i 
 \langle (\zeta^2 -1 )^i \rangle 
 \langle (\eta^2 -1 )^{k-i} \rangle \\\\
\ds  \langle (\eta^2 -1 )^{i} \rangle = 
\sum_{l=0}^i 
\binom{i}{l} (-1)^{l}
\langle \eta^{2(i-l)} \rangle 
 \end{array}
\label{m2}
\eea
In particular, the third moment or skewness, $H_3$,
which is responsible for the energy cascade, is
\bea
\ds H_3= \frac{  8  \psi^3 \left( \chi^3 - 1 \right) }
 {\left( 1+2  \psi^2 \left( 1+ \chi^2 \right) \right)^{3/2}  }
\label{H_3}
\eea 
For $\chi \ne$ 1, the skewness and all the odd order moments are different from zero,
and for $n>3$, all the absolute moments are rising functions of $R_{\lambda}$, 
thus $\Delta u_r$ exhibits an intermittency whose entity increases with the Reynolds number.

\begin{figure}
\vspace{-0.mm}
\centering
\includegraphics[width=0.45\textwidth]{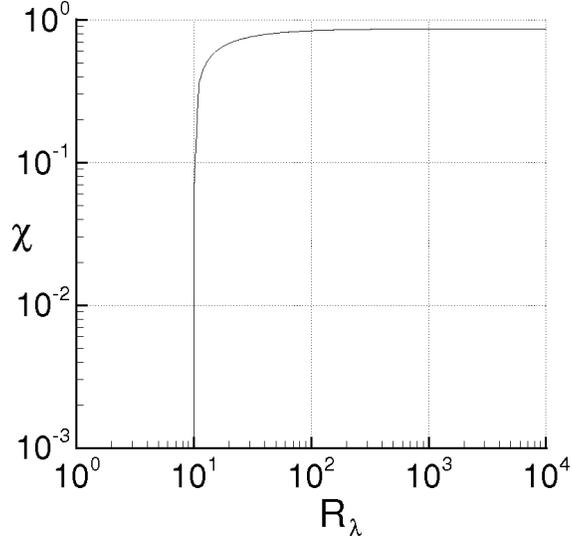}
\vspace{-2.mm}
\caption{Parameter $\chi$ plotted as the function of $R_{\lambda}$.}
\vspace{-0.mm}
\label{figura_2}
\end{figure}

All the statistical moments can be calculated once the function $\chi(R_\lambda)$
and the value of $\hat{\psi}_0$ are known.
The expression of $K(r)$ obtained in the first part of the work allows to
identify $H_3(0)$ and then fixes the relationship between $\psi_0$ and $\chi(R_\lambda)$
\bea
-H_3(0) = \frac{  8  {\psi_0}^3 \left(  1-\chi^3 \right) }
 {\left( 1+2  {\psi_0}^2 \left( 1+ \chi^2 \right) \right)^{3/2}  }
= \frac{3}{7}
\label{sk1}
\eea
where ${\psi}_0 = \psi(0, R_{\lambda})= O (\sqrt{R_{\lambda}})$ and 
$\chi = \chi(R_\lambda) > 0$.
This relationship does not admit solutions with $\chi > 0$ below a minimum value
of $(R_\lambda)_{min}$ dependent on $\hat{\psi}_0$.
According to the analysis of section \ref{s3} (Appendix B), $(R_\lambda)_{min}$
is chosen to 10.12, which corresponds to $\hat{\psi}_0 \simeq 1.075$.
(setting $\chi=0$, $R_\lambda$ = 10.12 in $H_3(0)$).
Varying the value of $(R_\lambda)_{min}$ from 8.5 to 15 would bring values of
$\hat{\psi}_0$ between 1.2 and 0.9, respectively.
In figure \ref{figura_2}, the function $\chi(R_\lambda)$ is shown for
 $\hat{\psi}_0 = 1.075$.
The limit  $\chi \simeq$ 0.86592 for $R_{\lambda} \rightarrow \infty$ is
reached independently of the value of $\hat{\psi}_0$.

The PDF of $\Delta u_r$ is expressed through the Frobenious-Perron equation
\bea
\begin{array}{l@{\hspace{+0.0cm}}l}
F(\Delta {u'}_r) = \hspace{-0.mm}
\ds \int_\xi \hspace{-0.mm}
\int_\eta  \hspace{-0.mm}
\int_\zeta \hspace{-0.mm}
p(\xi) p(\eta) p(\zeta) \
\delta \left( \Delta u'_r\hspace{-0.mm}-\hspace{-0.mm}\Delta {u}_r(\xi, \eta ,\zeta) \right)   
d \xi \ d \eta \ d \zeta
\end{array}
\label{frobenious_perron}
\eea 
where $\Delta {u}_r$ is calculated with Eq. (\ref{fluc4}), $\delta$ is the Dirac delta and $p$ is a gaussian PDF whose average value and standard deviation are equal to 0 and 1, respectively.

\bigskip

For non-isotropic turbulence or in more complex cases
with boundary conditions, the velocity spectrum could not satisfy the
Lindeberg condition, thus the velocity will be not distrubuted following 
a Gaussian PDF, and Eq. (\ref{fluc3}) changes its analytical form and can incorporate more intermittant terms \cite{Lehmann99} which give the deviation with respect to the isotropic turbulence.
Hence, the absolute statistical moments of $\Delta {u}_r$ will be greater than those calculated with Eq. (\ref{fluc4}), indicating that, in a more complex situation than the isotropic turbulence, the intermittency of $\Delta {u}_r$ can be significantly stronger.

\section{Results and discussion  \label{s8}}

In order to obtain informations about the validity of 
the proposed analysis, several results are now presented.

As the first result, the evolution in time of the correlation function is calculated with the proposed closure of the von K\'arm\'an-Howarth equation (Eq. (\ref{vk6})), where the boundary conditions are given by Eq. (\ref{bc}).
The turbulent kinetic energy and the spectrums $E(\kappa)$ and $T(\kappa)$ are calculated with Eq. (\ref{ke}) and Eqs. (\ref{Ek}), respectively.
The calculation is carried out for the initial Reynolds number of 
$Re = u(0) L_r/ \nu$ = 2000, where $L_r$ and $u(0)$ are, respectively, 
the characteristic dimension of the problem and the initial velocity standard deviation. The initial condition for the correlation function is 
$f(r) = \exp\left( -(r/\lambda_T)^2\right) $, where 
$\lambda_T/L_r$ = $1/(2 \sqrt{2})$, whereas $u(0)$ = 1. 
The dimensionless time of the problem is defined as $\bar{t} = t \ u(0)/L_r$.

Equation (\ref{vk}) was numerically solved adopting the Crank-Nicholson integrator scheme with variable time step, where the discretization of the space domain is made by $N-1$ intervals of the same amplitude $\Delta r$. 
This corresponds to a discretization of the Fourier space made by $N-1$ subsets in the interval $\left[ 0, \kappa_M \right]$, where $\kappa_M$ = $\pi/(2 \Delta r)$.
For the adopted initial Reynolds number, the choice $N$ = 1500, gives an adequate discretization, which provides $\Delta r < \ell$, for the whole simulation.
For what concerns $u$, it was calculated with Eq. (\ref{ke}) and the kinetic energy was checked to be equal to the integral over $\kappa$ of the energy spectrum. 
During the simulation, $T(\kappa)$ must identically satisfy Eq.(\ref{tk0})
(see Appendix A) which states that $T(\kappa)$ does not modify the kinetic energy.
According to the discretization of the Fourier space, the integral of $T(\kappa)$ 
is calculated with the trapezes rule from $0$ until to $\kappa_M$, at each time step,
therefore, the simulation will be considered to be accurate as long as
\bea
\int_0^{\kappa_M} T(\kappa) d \kappa \simeq \int_0^{\infty}
 T(\kappa) d \kappa = 0
\label{tk0a}
\eea   
namely, when $T(\kappa) \simeq 0$ for $\kappa > \kappa_M$.
As the simulation advances, according to Eq. (\ref{vk6}), the energy cascade
determines variations of $E(\kappa)$ and $T(\kappa)$ for wave-numbers
whose values rise with the time, then Eq. (\ref{tk0a}) holds until to a certain time,
where these wave-number are about equal to $\kappa_M$.
At higher times, the variations of $T(\kappa)$ can occur for $\kappa > \kappa_M$, out of the interval $(0,  \kappa_M)$, thus Eq. (\ref{tk0a}) could be not satisfied.
For this reason, the simulation is stopped as soon as the
following condition is achieved \cite{NAG}
\bea
\ds \vert \int_0^{\kappa_M} T(\kappa) d \kappa \vert >
 \frac{1}{N^2} \int_0^{\kappa_M} \vert T(\kappa) \vert  d \kappa 
\eea
At the end of several simulations, we obtain $\Delta r \approx 0.8 \ \ell$,
and, in this situation, the energy spectrum is here supposed to be fully developed.
\begin{figure}
\vspace{-0.mm}
	\centering
\includegraphics[width=0.45\textwidth]{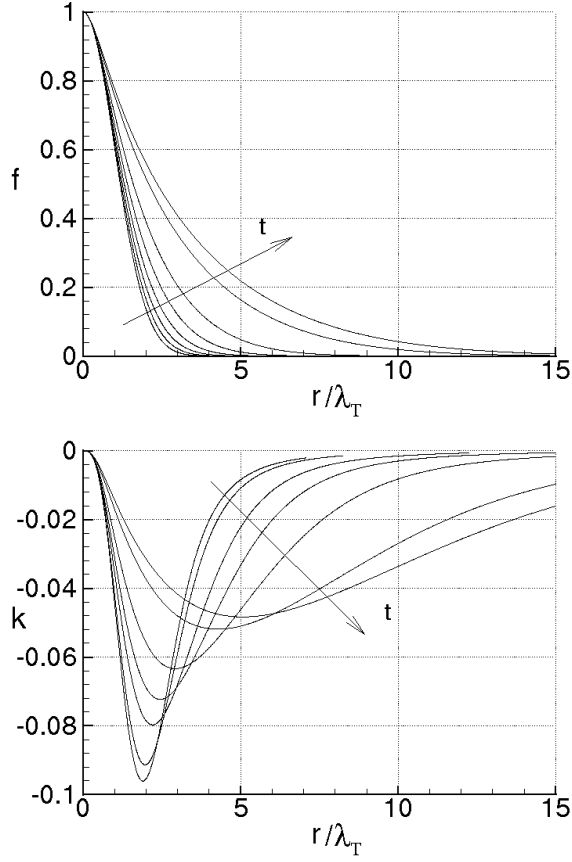}
\vspace{-0.mm}
\caption{Correlation functions, $f$ and $k$ versus the separation distance at the times of simulation $\bar{t}$ = 0, 0.1, 0.2, 0.3, 0.4, 0.5, 0.6, 0.63.}
\vspace{-0.mm}
\label{figura_6}
% f_k.jpg: 100dpi, width=7.62cm, height=10.80cm, bb=0 0 300 425
\end{figure}

The diagrams of Fig. \ref{figura_6} show the correlation functions $f(r)$ and $k(r)$ vs. the dimensionless distance $r/\lambda_T$, at different times of simulation. 
The kinetic energy and Taylor scale vary according to Eqs. (\ref{vk6}) and (\ref{ke}), thus $f(r)$ and $k(r)$ change in such a way that the length scales 
associated to their variations diminish as the time increases, whereas the maximum of $\vert k \vert$ decreases. 
At the final instants of the simulation, one obtains that $f - 1 =$ O( $r^{2/3}$) 
for $r/\lambda_T=$ O(1), whereas the maximum of $\vert k \vert$ is about 0.05.
These results are in very good agreement with the numerous data of the literature \cite{Batchelor53} which concern the evolution of correlation functions.
\begin{figure}
\vspace{-0.mm}
	\centering
\hspace{-0.mm}
\includegraphics[width=0.49\textwidth]{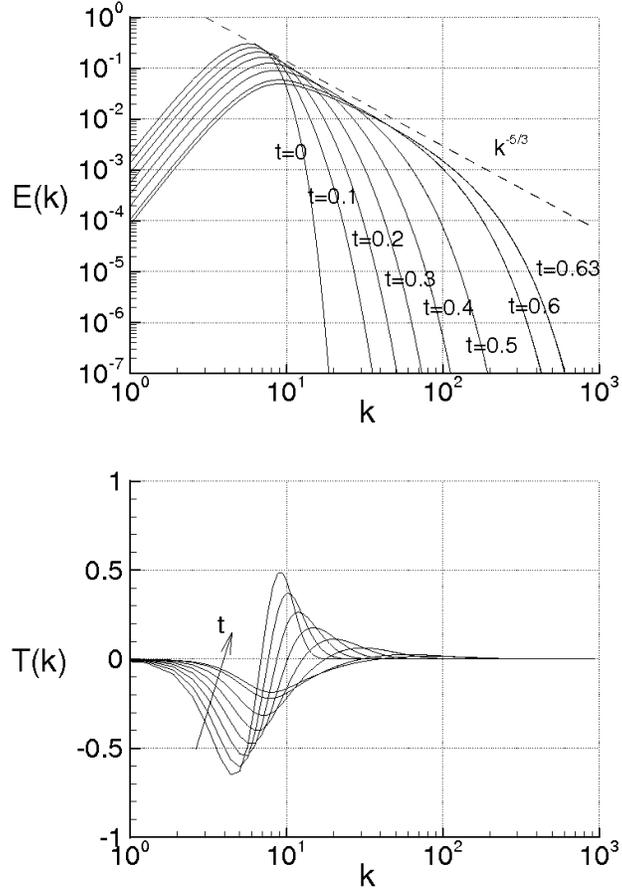}
\vspace{-0.mm}
\caption{Plot of $E(\kappa)$ and $T(\kappa)$ at the diverse times of simulation.}
% Ek_Tk.jpg: 100dpi, width=10.16cm, height=14.40cm, bb=0 0 400 567
\vspace{-0.mm}
\label{figura_7}
\end{figure}
Figure \ref{figura_7} shows the diagrams of $E(\kappa)$ and $T(\kappa)$ 
for the same times, where the dashed line in the plot of $E(\kappa)$, 
represents the $-5/3$ Kolmogorov law \cite{Kolmogorov41}.
\begin{figure}
\vspace{+0.mm}
	\centering
\hspace{-0.mm}
\includegraphics[width=0.42\textwidth]{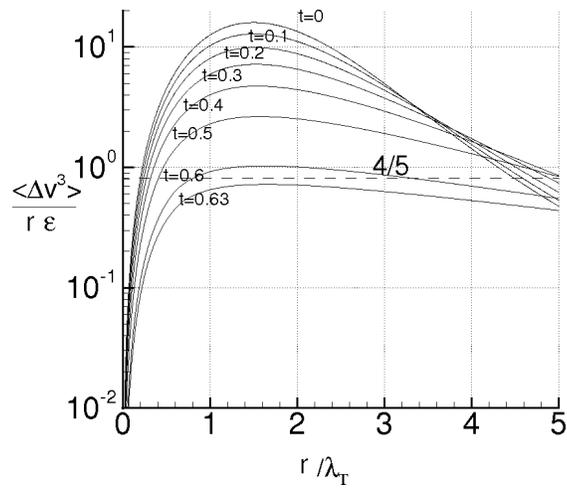}
\vspace{-0.mm}
\caption{The Kolmogorov function versus $r/\lambda_T$ for different times of simulation. The dashed line indicates  the value  4/5.}
\vspace{-0.mm}
% f_kolm.jpg: 100dpi, width=10.16cm, height=7.16cm, bb=0 0 400 282
\label{figura_8}
\end{figure}
The spectrums $E(\kappa)$ and $T(\kappa)$ vary with time according to Eqs. (\ref{vk6}) and (\ref{Ek}) and depend on the initial condition.
At the end of simulation, the energy spectrum $E(\kappa)$ can be compared with the dashed line in an opportune interval of wave-numbers.
This arises from the developed correlation function, which behaves like
$f -1$ = O ($r^{2/3}$) for $r = O(\lambda_T)$. 
\begin{figure}
\vspace{-0.mm}
	\centering
\hspace{+10.mm}
\includegraphics[width=0.45\textwidth]{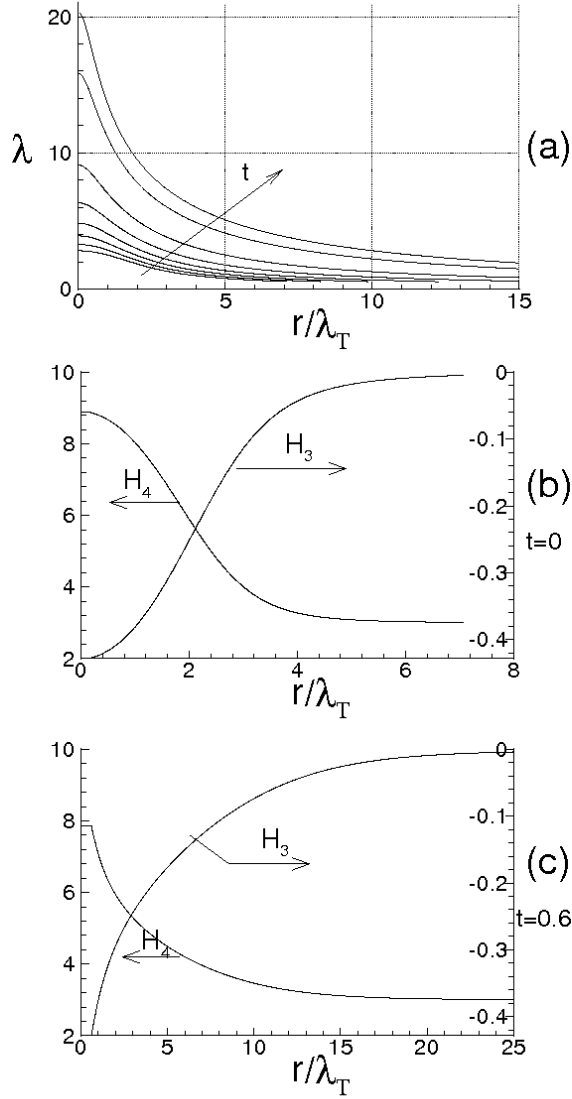}
% lambda_H.jpg: 100dpi, width=10.16cm, height=14.40cm, bb=0 0 400 567
\vspace{-0.mm}
\caption{(a) Maximum finite size Lyapunov exponent at the times of simulation $\bar{t}$ = 0, 0.1, 0.2, 0.3, 0.4, 0.5, 0.6, 0.63; (b) and (c) skewness and Flatness versus $r/\lambda_T$ at t = 0 and t = 0.6, respectively.}
\vspace{-0.mm}
\label{figura_9}
\end{figure}

Next, the Kolmogorov function $Q(r)$ and Kolmogorov constant $C$, are determined with
the proposed analysis, using the previous results of the simulation. 

Following the Kolmogorov theory, the Kolmogorov function, which is defined as  
\bea 
\ds Q(r) = - \frac{\langle (\Delta u_r)^3 \rangle} { r \varepsilon}
\label{k_f}
\eea
is constant with respect to $r$, and is equal to 4/5 as long as $r/\lambda_T = O(1)$. 
As shown in Fig. \ref{figura_8}, for $\bar{t} =0$, the maximum of $Q(r)$ is much greater than 4/5 and its variations with $r/\lambda_T$ can not be neglected.
This is the consequence of the choice of the initial correlation function.
At the successive times, the maximum of $Q(r)$ decreases until to the final instants, where, with the exception of $r/\lambda_T \approx 0$, $Q(r)$ exhibits variations which are less than those calculated at the previous times in a wide range of $r/\lambda_T$, with a maximum which can be compared to 0.8.

The Kolmogorov constant $C$ is also calculated by definition
\bea
E(\kappa) = C \frac{\varepsilon^{2/3} } {\kappa^{5/3}}
\label{k_c}
\eea 
This is here determined, as the value of $C$ which makes the curve represented by  
Eq. (\ref{k_c}) to be tangent to the energy spectrum $E(\kappa)$ 
previously calculated.
At end simulation, $C \simeq $ 1.932, namely $C$ and $Q_{max}$ agree with the corresponding quantities known from the literature.

For the same simulation, Fig. \ref{figura_9}a shows the maximal finite scale Lyapunov exponent, calculated with Eq. (\ref{lC}), where $\lambda$ 
varies according to $f$.
For $t = 0$, the variations of $\lambda$ are the result of the adopted initial
correlation function which is a gaussian, whereas as the time increases, the
variations of $f$ determine sizable increments of $\lambda$ and of its slope in
proximity of the origin. 
Then, for developed spectrum, since $f -1$ = O($r^{2/3}$), 
the maximal finite scale Lyapunov exponent behaves like $\lambda \approx r^{-2/3}$.
Thus, the diffusivity coefficient associated to the relative motion between two fluid particles,  defined as $D(r) \propto \lambda r^2$, here satisfies the famous
Richardson scaling law $D(r) \approx r^{4/3}$\cite{Richardson26}.
\begin{figure}
\vspace{-0.mm}
	\centering
\includegraphics[width=0.48\textwidth]{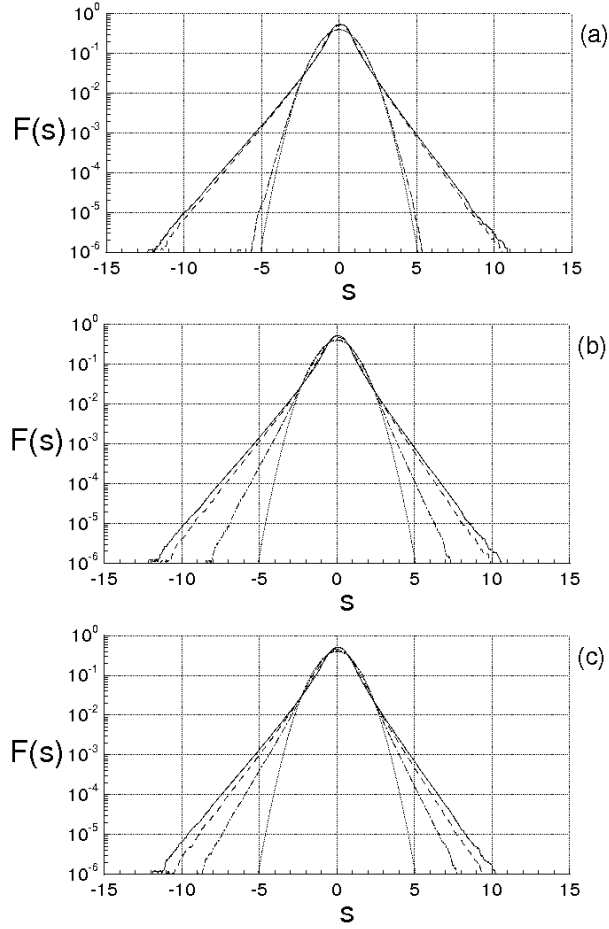}
% pdf_dv_abc.jpg: 100dpi, width=11.43cm, height=16.18cm, bb=0 0 450 637
\vspace{-0.mm}
\caption{PDF of the velocity difference fluctuations at the times $\bar{t}=$0 (a), $\bar{t}=$ 0.5 (b) and $\bar{t}=$0.6 (c). 
Continuous lines are for $r=$0, dashed lines are for $r/\lambda_T =$1, 
dot-dashed lines are for  $r/\lambda_T =$5, dotted lines are for gaussian PDF.}
\vspace{-0.mm}
\label{figura_10}
\end{figure}

In the diagrams of Figs. \ref{figura_9}b and \ref{figura_9}c, skewness and flatness of $\Delta u_r$ are shown in terms of $r$ for $\bar{t}$ = 0 and 0.6.
The skewness, $H_3$ is first calculated with Eq. (\ref{H_3_01}), 
then $H_4$ has been determined using Eq. (\ref{m1}). 
At $\bar{t}=0$, $\vert H_3 \vert$ starts from 3/7 at the origin with small slope, then decreases until to reach small values. $H_4$ also exhibits small derivatives near the origin, where $H_4\gg$ 3, thereafter it decreases more rapidly than $\vert H_3 \vert$. 
At $\bar{t} = $0.6, the diagram importantly changes and exhibits different shapes.
The Taylor scale and the corresponding Reynolds number are both changed, so that the variations of $H_3$ and $H_4$ are associated to smaller distances, whereas the 
flatness at the origin is slightly less than that at $t =0$. Nevertheless, these variations correspond to higher $r/\lambda_T$ than those for $t$ = 0, and also in this case, $H_4$ reaches the value of 3 more rapidly than $H_3$ tends to zero.

The PDFs of $\Delta u_r$ are calculated with Eqs. (\ref{frobenious_perron})
and (\ref{fluc4}), and are shown in Fig. \ref{figura_10} in terms of the dimensionless abscissa 
\bea
\ds s = \frac{\Delta u_r} 
{ \langle (\Delta u_r)^2 \rangle^{1/2}  }
\nonumber
\eea
where, these distribution functions are normalized, in order that their standard 
deviations are equal to the unity.
The figure  represents the distribution functions of $s$ for several
$r/\lambda_T$, at $\bar{t}$ = 0, 0.5 and 0.6, where the dotted curves represent the gaussian distribution functions.
The calculation of $H_3(r)$ is first carried out with Eq. (\ref{H_3_01}), then the function $\psi(r, R_\lambda)$ is identified through Eq. (\ref{H_3}), 
and finally the PDF is obtained with  Eq. (\ref{frobenious_perron}).
For $t$ = 0 (see Fig. \ref{figura_10}a) and according to the evolutions of $H_3$ and $H_4$, the PDFs calculated at $r/\lambda_T=$ 0 and 1, are quite similar each other, whereas for $r/\lambda_T=$ 5, the PDF is almost a gaussian function.
Toward the end of the simulation, (see Fig. \ref{figura_10}b and c), the two PDFs
calculated at $r/\lambda_T=$ 0 and 1, exhibit more sizable differences, whereas for $r/\lambda_T=$ 5, the PDF differs very much from a gaussian PDF. 
This is in line with the plots of $H_3(r)$ and $H_4(r)$ of Fig. \ref{figura_9}.
\begin{figure}
\vspace{-0.mm}
	\centering
\hspace{-0.mm}
\includegraphics[width=0.45\textwidth]{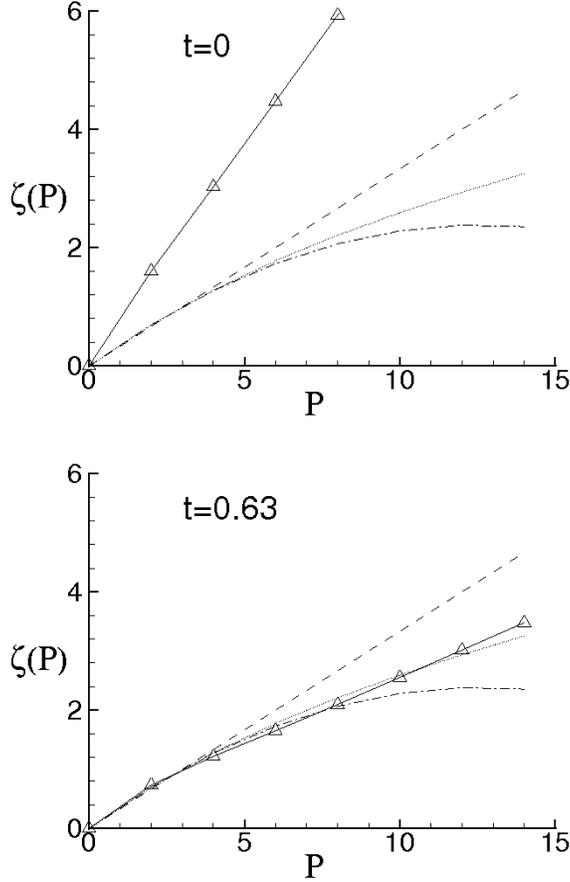}
% zp.jpg: 100dpi, width=7.62cm, height=10.80cm, bb=0 0 300 425
\vspace{-0.mm}
\caption{Scaling exponents of longitudinal velocity difference versus the order moment at different times. Continuous lines with solid symbols are for the present data. Dashed lines are for Kolmogorov K41 data \cite{Kolmogorov41}. Dashdotted lines are for Kolmogorov K62 data \cite{Kolmogorov62}.
 Dotted lines are for She-Leveque data \cite{She-Leveque94}}
\vspace{-0.mm}
\label{figura_11}
\end{figure}

\bigskip

Next, the spatial structure of $\Delta u_r$, given by Eq. (\ref{fluc4}), is analyzed 
using the previous results of the simulation. 
According to the various works \cite{Kolmogorov62, She-Leveque94, Benzi91}, 
$\Delta u_r$ behaves quite similarly to a multifractal system, where $\Delta u_r$ 
obeys to a law of the kind 
$
\Delta u_r(r) \approx r^q
$
where the exponent $q$ is a fluctuating function of space.
This implies that the statistical moments of $\Delta u_r(r)$ are expressed through 
different scaling exponents $\zeta(P)$ whose values depend on the moment order $P$, i.e.
\bea
\left\langle (\Delta u_r)^{P}(r) \right\rangle  = A_P \ r^{\zeta(P)}
\label{fractal}
\eea
\begin{table}[b] 
  \begin{center} 
  \begin{tabular}{lrrrrrrrrrrrrrr}
\hline
P      \    & 1  \   & 2   \  & 3  \   & 4   \ & 5  \  & 6   \ & 7  \  & 8  \  & 9   \ & 10 \  & 11 \  & 12  \ & 13  \ & 14   \\
$\zeta$(P) \ & 0.36 \ & 0.71 \ & 0.99 \ & 1.19 \ & 1.41 \ & 1.61 \ & 1.84 \ & 2.04 \ & 2.25 \ & 2.49 \ & 2.72 \ & 2.93 \ & 3.15 \ & 3.38 \\
\hline
 \end{tabular}
\caption{Scaling exponents of the longitudinal velocity difference.}
  \end{center} 
  \vspace{-5.mm}
\end{table} 
These scaling exponents are here identified through a best fitting procedure, 
in the intervals ($a_P, a_P$ $ + \lambda_T$), where the endpoints $a_P$ are unknown 
quantities which have to be determined.
The location of these intervals depends on $P$ and varies with the time. The calculation of the endpoints $a_P$ and of $\zeta_P$ and $A_P$ is carried out through a minimum square method which for each moment order is applied to the following optimization problem
\bea
\ds J_P(a_P, \zeta_P, A_P) \hspace{-1.mm} \equiv 
%\ds \frac{1}{\vert b_P -a_P \vert^{2 \zeta(P)+1}} 
\int_{a_P}^{a_P +\lambda_T} 
\ds ( \langle (\Delta u_r)^P \rangle - A_P r^{\zeta(P)} )^2 dr  = \mbox{min}, \  
 P = 1, 2, ...
\eea
where $(\langle \Delta u_r^P)\rangle$ are calculated with Eqs. (\ref{m1}).
\\
Figure \ref{figura_11} shows the comparison between the scaling exponents here obtained  (continuous lines with solid symbols) and those of the Kolmogorov theories K41 \cite{Kolmogorov41} (dashed lines) and K62 \cite{Kolmogorov62} (dashdotted lines), and those given by She-Leveque \cite{She-Leveque94} (dotted curves).
At $t=$ 0, the values of $\zeta(P)$ are the result of the chosen initial condition.
As the time increases, the correlation function changes causing variations in the statistical
moments of $\Delta u_r(r)$. As result, $\zeta(P)$ gradually diminish and exhibit a variable slope which depends on the moment order $P$, until to reach the situation of Fig. \ref{figura_11}b, where the simulation is just ended. 
The dimensionless moments of $\Delta u_r(r)$ are changed. The plot of $\zeta(P)$ shows that near the origin, $\zeta(P) \simeq P/3$, and that the values of $\zeta(P)$ seem to be in agreement with the those proposed by She-Leveque. More in detail, Table I reports these scaling exponents in terms of the moments order, calculated for $\bar{t} = 0.63$.
These values are the consequence of the spatial variations of the skewness, calculated using Eq. (\ref{H_3_01}), and of the quadratic terms due to the inertia and pressure forces into the expression of the velocity difference, which make $\langle (\Delta u_r)^P \rangle$ a quantity quite similar to a multifractal system.

Other simulations with different initial correlation functions and Reynolds
numbers have been carried out, and all of them lead to analogous results,
in the sense that, at the end of the simulations, the diverse quantities 
such as $Q(r)$, $C$ and $\zeta(P)$ are quite similar to those just calculated.
For what concerns the effect of the Reynolds number, its increment determines
a wider range of the wave-numbers where $E(\kappa)$ is comparable with the Kolmogorov 
law and a smaller dissipation energy rate in accordance to Eq. (\ref{ke}).
\bigskip
\begin{table}[b] 
  \begin{center} 
  \begin{tabular}{lrrrr} 
\hline
Moment \ & $R_\lambda \approx 10$ \ & $R_\lambda=10^2$ \ & $R_\lambda=10^3$ \ & Gaussian\\[2pt] 
Order    &  P. R.                 & P. R.            &  P. R. \           & Moment \\[2pt] 
\hline
3        & -.428571               & -.428571         & -.428571          & 0      \\
4        &   3.96973              &  7.69530         & 8.95525           & 3      \\
5        & -7.21043               &  -11.7922        & -12.7656          & 0      \\
6        &  42.4092               &  173.992         & 228.486           & 15     \\
7        & -170.850               &  -551.972        & -667.237          & 0      \\
8        &  1035.22               &  7968.33         & 11648.2           & 105    \\
9        &  -6329.64              &  -41477.9        & -56151.4          & 0      \\
10       & 45632.5                &  617583.         & 997938.           & 945    \\
\hline
 \end{tabular}
\caption{Dimensionless statistical moments of $F(\partial u_r/\partial r)$ at different
Taylor scale Reynolds numbers. P.R. as for ''present results''.}
  \end{center} 
  \vspace{-5.mm}
\end{table} 

\begin{figure}[t]
\vspace{-0.mm}
	\centering
\includegraphics[width=0.45\textwidth]{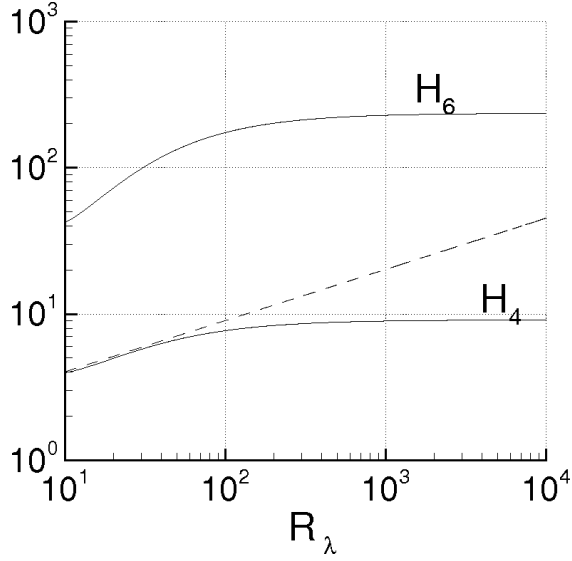}
\vspace{-0.mm}
\caption{Dimensionless moments $H_4(0)$ and $H_6(0)$ plotted vs. $R_{\lambda}$.
Continuous lines are for the present results. 
The dashed line is the tangent to the curve of $H_4(0)$ in $R_\lambda \approx$ 10.}
\vspace{-0.mm}
\label{figura_3}
% H_Re.jpg: 100dpi, width=8.89cm, height=6.27cm, bb=0 0 350 247
\end{figure}

In order to study the evolution of the intermittency vs. the Reynolds number,
Table II gives the first ten statistical moments of $F(\partial u_r/\partial r)$. 
These are calculated with Eqs. (\ref{m1}) and (\ref{m2}), for $R_{\lambda}$ = 10.12, 100 and 1000, and are shown in comparison with those of a gaussian distribution function.
It is apparent that a constant nonzero skewness of the longitudinal velocity derivative, causes an intermittency which rises with $R_\lambda$ 
(see Eq. (\ref{fluc4})).
More specifically, Fig. \ref{figura_3} shows the variations of $H_4(0)$ and $H_6(0)$
(continuous lines) in terms of $R_\lambda$, calculated with Eqs. (\ref{m1}) 
and (\ref{m2}), with $H_3(0) = -3/7$.
These moments are rising functions of $R_{\lambda}$ for 
10 $\lesssim R_{\lambda} \lesssim$ 700, whereas for higher $R_{\lambda}$ these tend to
the saturation and such behavior also happens for the other absolute moments.
According to Eq. (\ref{m1}), in the interval 10 $ \lesssim R_{\lambda} \lesssim$  70, 
$H_4$ and $H_6$ result to be about proportional to $R_{\lambda}^{0.34}$ and $R_{\lambda}^{0.78}$,
respectively, and the intermittency increases with the Reynolds number until to 
$R_{\lambda} \approx$ 700,  where it ceases to rise so quickly. 
\begin{figure}[t]
\vspace{-0.mm}
	\centering
\includegraphics[width=0.48\textwidth]{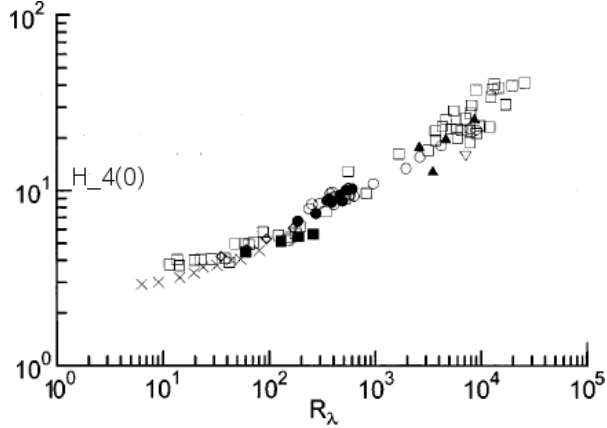}
\vspace{-0.mm}
\caption{Flatness $H_4(0)$ vs. $R_{\lambda}$. These data are from  Ref.\cite{Antonia97}.}
\vspace{-0.mm}
\label{antonia}
% H_Re.jpg: 100dpi, width=8.89cm, height=6.27cm, bb=0 0 350 247
\end{figure}
\begin{figure}[t]
\vspace{-0.mm}
	\centering
\includegraphics[width=0.50\textwidth]{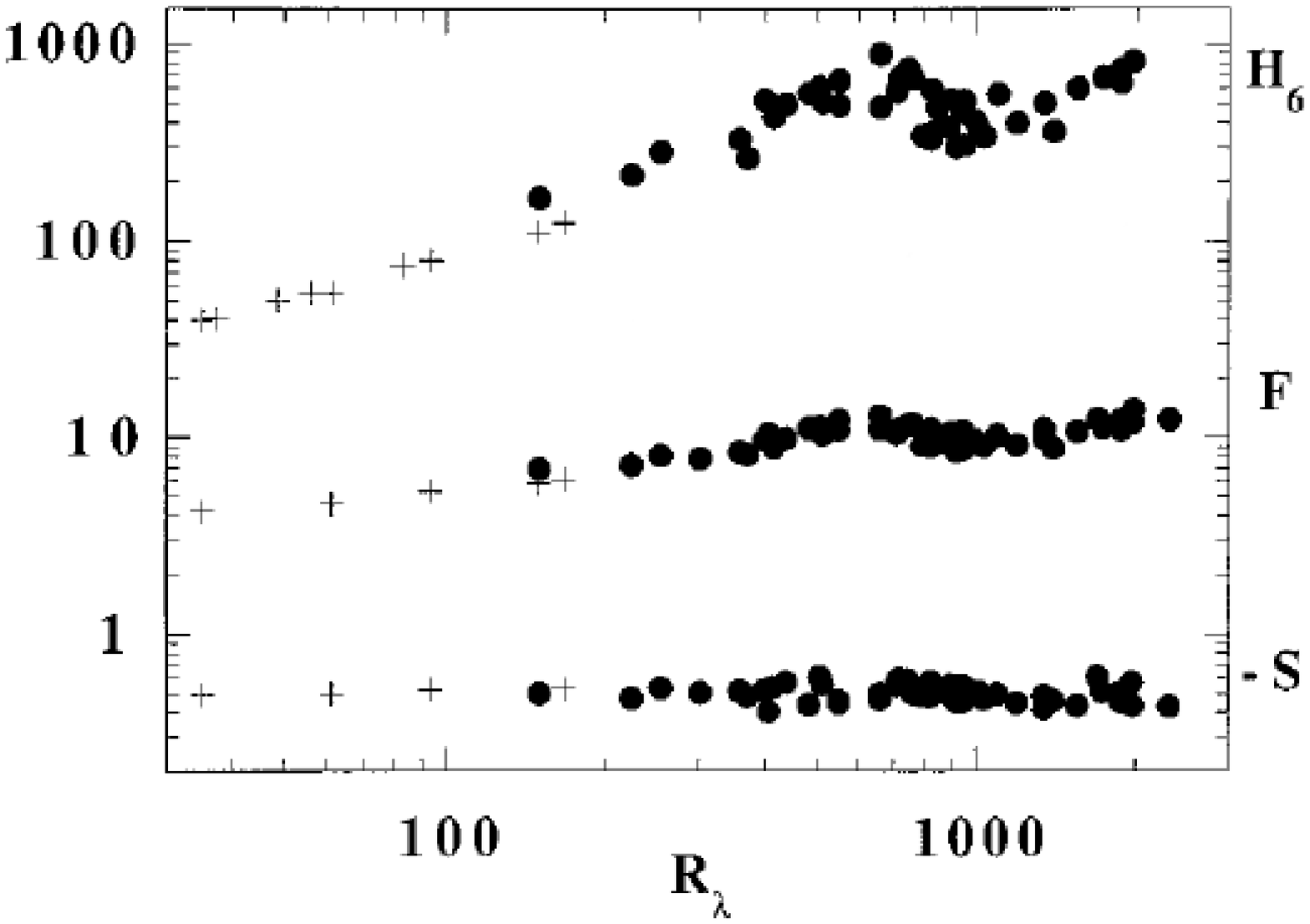}
\vspace{-0.mm}
\caption{Skewness $S = H_3(0)$, Flatness $F =H_4(0)$ and hyperflatness $H_6(0)$ vs. $R_{\lambda}$. These data are from  Ref.\cite{Tabeling97}.}
\vspace{-0.mm}
\label{tabeling0}
% H_Re.jpg: 100dpi, width=8.89cm, height=6.27cm, bb=0 0 350 247
\end{figure}
This behavior, represented by the continuous lines, depends on the fact that 
$\psi \approx \sqrt{R_\lambda}$, and results to be in very good agreement with the data 
of Pullin and Saffman \cite{Pullin93}, for 10 $\lesssim R_{\lambda} \lesssim$  100.
\begin{figure}[t]
\vspace{+0.mm}
\centering
\hspace{0.mm}
\includegraphics[width=0.45\textwidth]{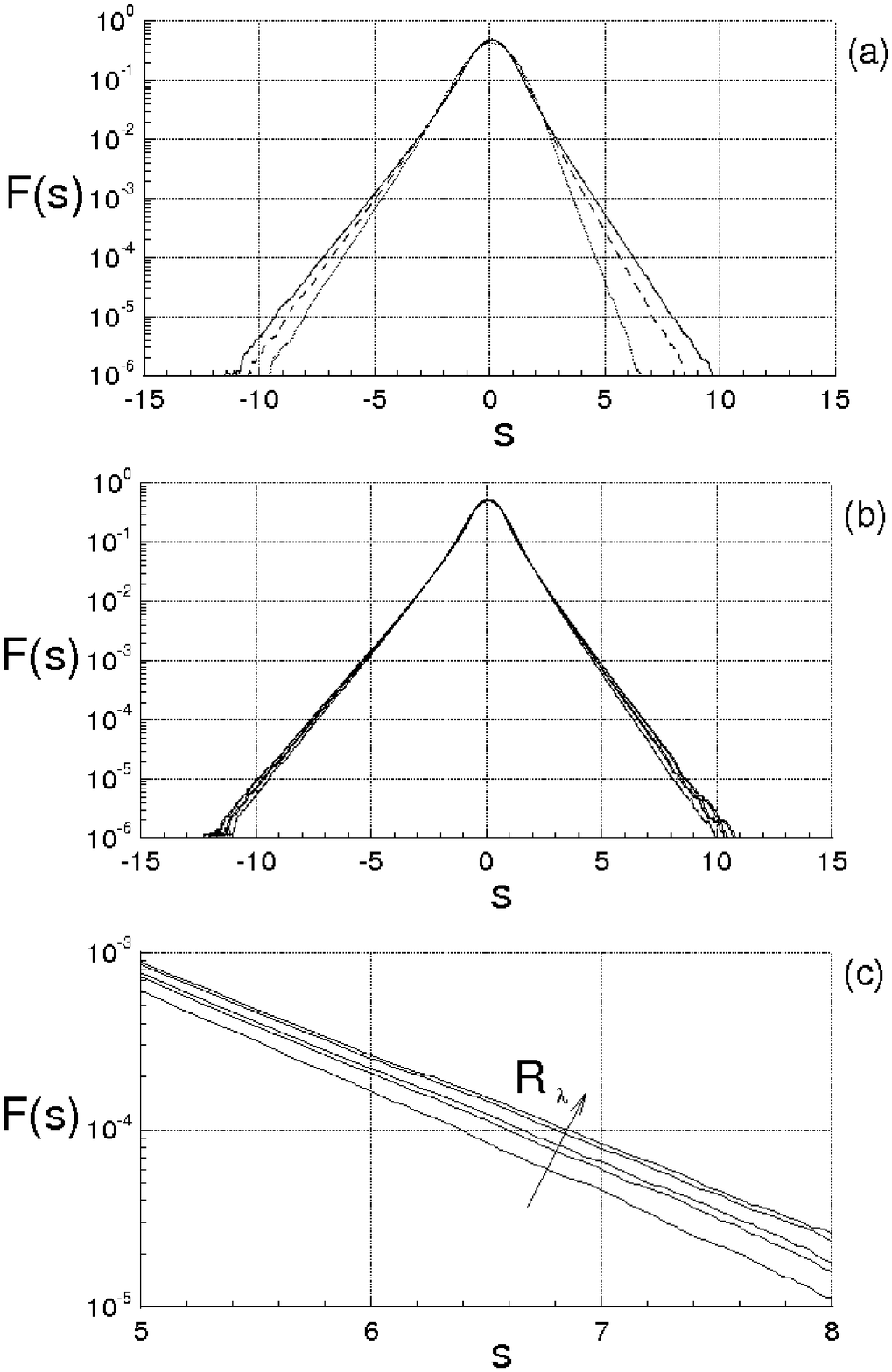}
\vspace{-0.mm}
\caption{Log linear plot of the PDF of $\partial u_r/\partial r$ for different
$R_\lambda$. (a): dotted, dashdotted and continuous lines are
for $R_\lambda$ = 15, 30 and 60, respectively. (b) and (c) PDFs for 
$R_{\lambda}$ = 255, 416, 514, 1035 and 1553. (c) represents an enlarged part
of the diagram (b) }
\vspace{-0.mm}
\label{figura_4}
% pdf_abc.jpg: 100dpi, width=10.16cm, height=14.40cm, bb=0 0 400 567
\end{figure}
\begin{figure}[t]
\vspace{-0.mm}
\centering
\hspace{0.mm}
\includegraphics[width=0.42\textwidth]{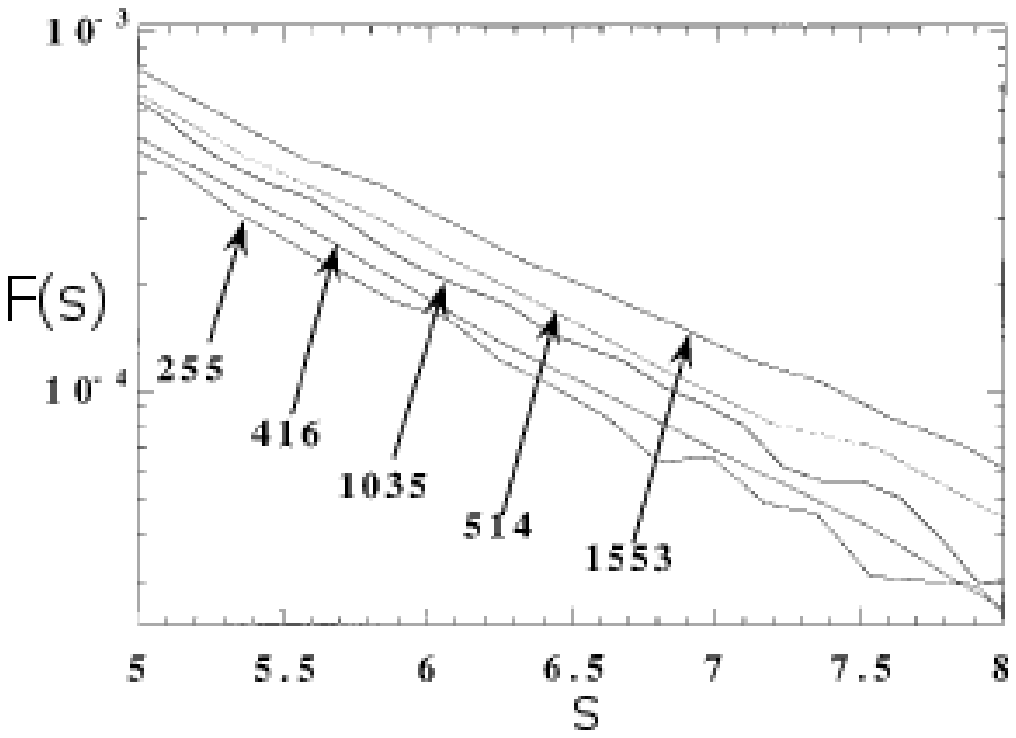}
\vspace{-0.mm}
\caption{PDF of $\partial u_r/\partial r$ for 
$R_{\lambda}$ = 255, 416, 514, 1035 and 1553. These data are from Ref. \cite{Tabeling97} }
\vspace{-0.mm}
\label{tabeling}
% pdf_abc.jpg: 100dpi, width=10.16cm, height=14.40cm, bb=0 0 400 567
\end{figure}
Figure \ref{figura_3} can be compared with the data collected by Sreenivasan and Antonia \cite{Antonia97}, which are here reported into Fig. \ref{antonia}.
These latter are referred to several measurements and simulations obtained in different situations which can be very far from the isotropy and homogeneity conditions. 
Nevertheless a comparison between the present results and those of 
Ref. \cite{Antonia97} is an opportunity to state if the two data exhibit 
elements in common.
According to  Ref. \cite{Antonia97}, the flatness monotonically rises with $R_\lambda$ with a rising rate which agrees with Eq. (\ref{m2}) for 
$10 \lesssim R_\lambda \lesssim 60$ (dashed line, Fig. \ref{figura_3}), whereas the skewness seems to exhibit minor variations.
Thereafter, $H_4$ continues to rise with about the same rate, 
without the saturation observed in Fig. \ref{figura_3}.
The weaker intermittency calculated with the present analysis arise from the isotropy which makes the velocity fluctuation a gaussian random variable, while,
as seen in sec. \ref{s4}, without the isotropy condition, the flatness of velocity 
and of velocity difference can be much greater than that of the isotropic case.

Again, the obtained results are compared with the data of Tabeling {\it et al} 
\cite{Tabeling96,  Tabeling97}, where, in an experiment using low temperature helium gas  between two counter-rotating cylinders (closed cell), the authors measure the PDF of $\partial u_r/\partial r$ and its moments.
Also in this case the flow can be quite far from to the isotropy condition.
In fact, these experiments pertain wall-bounded flows, where the walls could importantly influence the fluid velocity in proximity of the probe. 
The authors found that the higher moments than the third order, first increase with
$R_{\lambda}$ until to $R_{\lambda} \approx$ 700, then exhibit a lightly non-monotonic evolution with respect to $R_{\lambda}$, and finally cease their
 variations denoting a transition behavior (See Fig. \ref{tabeling0}). 
As far as the skewness is concerned, the authors observe small percentage variations. 
Although the isotropy does not describe the non-monotonic evolution near 
$R_{\lambda} =$ 700, the results obtained with Eq. (\ref{fluc4}) can be considered comparable with those of Refs. \cite{Tabeling96, Tabeling97}, resulting also
in this case, that the proposed analysis gives a weaker intermittency with respect to 
Refs. \cite{Tabeling96, Tabeling97}.

The normalized PDFs of $\partial u_r/\partial r$ are calculated with Eqs. (\ref{frobenious_perron}) and (\ref{fluc4}), and are shown in Fig. \ref{figura_4} 
in terms of the variable $s$, which is defined as  
\bea
\ds s = \frac{\partial u_r/\partial r} 
{ \left\langle (\partial u_r/\partial r)^2\right\rangle^{1/2}  }
\nonumber
\eea
Figure \ref{figura_4}a shows the diagrams for $R_{\lambda} =$ 15, 30 and 60, where
the PDFs vary in such a way that $H_3(0) = -3/7$.
\\
As well as in Ref. \cite{Tabeling97}, Figs. 4b and 4c give the PDF for
$R_{\lambda}$ = 255, 416, 514, 1035 and 1553,  where these last Reynolds numbers are calculated through the  Kolmogorov function given in Ref. \cite{Tabeling97}, with 
$H_3(0) = -3/7$.
In particular, Fig. \ref{figura_4}c represents the enlarged region of Fig. \ref{figura_4}b,  where the tails of PDF are shown for $5 < s < 8$.
According to Eq. (\ref{fluc4}), the tails of the PDF rise in the interval  
10 $\lesssim R_{\lambda} \lesssim$ 700, whereas at higher  $R_{\lambda}$, smaller variations occur.
Although the non-monotonic trend observed in Ref. \cite{Tabeling97}, 
Fig. \ref{figura_4}c shows that the values of the PDFs
calculated with the proposed analysis, for $5 < s < 8$, exhibit the same
order of magnitude of those obtained by Tabeling {\it et al} 
\cite{Tabeling97} which are here shown in Fig. \ref{tabeling}.
\begin{figure}[t]
\vspace{0.mm}
 \centering
\hspace{-0.mm}
\includegraphics[width=0.50\textwidth]{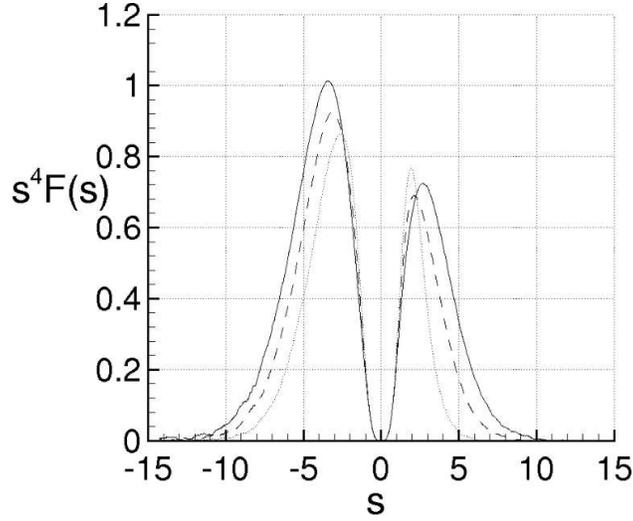}
\vspace{-0.mm}
\caption{Plot of the integrand $s^4 F(s)$ for different
$R_\lambda$. Dotted, dashdotted and continuous lines are
for $R_\lambda$ = 15, 30 and 60, respectively.}
\vspace{-0.mm}
\label{figura_5}
% dv4_f.jpg: 100dpi, width=10.16cm, height=7.16cm, bb=0 0 400 282
\end{figure}

Asymmetry and intermittency of the distribution functions are also 
represented through the integrand function of the $4^{th}$
order moment of PDF, which is
$
J_4(s) = s^4 F(s)
\nonumber
$.
This function is shown in terms of $s$, in Fig. \ref{figura_5}, for
 $R_\lambda$ = 15, 30 and 60.

\section{\bf  Conclusions  \label{s9}}

The proposed analysis is based on the conjecture which states that  
the turbulence is caused by the bifurcations of the velocity field. 
The main limitation of this analysis is that it only studies 
the developed homogeneous-isotropic turbulence,
whereas this does not consider the intermediate stages of the turbulence.

The results of this analysis can be here summarized:

\begin{enumerate}

\item The Lyapunov analysis of the relative motion provides
an explanation of the physical mechanism of the energy cascade in turbulence
and gives a closure of the von K\'arm\'an-Howarth equation. 

The fluid incompressibility determines that the inertia forces transfer the kinetic energy between the length scales without changing the total kinetic energy.
This implies that the skewness of the longitudinal velocity derivative is 
a constant of the present analysis and that the energy cascade mechanism does
not depend on the Reynolds number.

\item The momentum equations written using the referential coordinates allow
to factorize the velocity fluctuation and to express it in Lyapunov exponential form
of the local fluid deformation.
The statistics of $\Delta u_r$ can be inferred looking at the Fourier series of the velocity difference. This is a non-Gaussian statistics, where the constant skewness of 
$\partial u_r/ \partial r$ implies that the other higher absolute moments 
increase with the Taylor-scale Reynolds number.

\item The momentum equations written using the referential coordinates allow 
the velocity fluctuation to be expressed by means of the
Lyapunov analysis of the kinematics of local fluid deformation.
The Fourier series of the velocity difference provides the statistics of
$\Delta u_r$. This is a non-Gaussian statistics, where the constant skewness of 
$\partial u_r/ \partial r$ implies that the other higher absolute moments 
increase with the Taylor-scale Reynolds number. 

\item The closure of the von K\'arm\'an-Howarth equation, shows that the mechanism of energy cascade gives energy spectrums that can be compared with the Kolmogorov law $\kappa^{-5/3}$ in an opportune range of wave-numbers.

\item For developed energy spectrums, the Kolmogorov function exhibits, in an opportune range of $r$, small variations much less than at the previous times, and its maximum is quite close to 4/5, whereas the Kolmogorov constant is about equal to 1.93. 
As the consequence, the maximal finite scale Lyapunov exponent and the diffusivity coefficient vary according to the Richardson law when the separation distance is of the order of the Taylor scale.

\item The analysis also determines the scaling exponents of the moments of the
      longitudinal velocity difference through a best fitting procedure. 
      For developed energy spectrum, these exponents show variations with the moment order
      which seem to be consistent with those known from the literature.
\end{enumerate}

\section{\bf  Acknowledgments}

This work was partially supported by the Italian Ministry for the 
Universities and Scientific and Technological Research (MIUR).
The author is indebted with the referees and the Editor for their useful comments.

\section{\bf Appendix A}

The von K\'arm\'an-Howarth equation gives the evolution in time of the
longitudinal correlation function for isotropic turbulence.
The correlation function of the velocity components is the symmetrical 
second order tensor
$
\ds R_{i j} ({\bf r}) = \left\langle u_i u_j' \right\rangle 
$, 
where $u_i$ and  $u_j'$ are the velocity components at ${\bf x}$ and 
${\bf x} + {\bf r}$, respectively, being $\bf r$ the separation vector.
The equations for $R_{i j}$ are obtained by the Navier-Stokes equations 
written in the two points ${\bf x}$ and ${\bf x} + {\bf r}$ \cite{Karman38, Batchelor53}.
For isotropic turbulence $R_{i j}$ can be expressed as
\bea
R_{i j} ({\bf r}) = u^2 \left[ (f -g) \frac{r_i r_j}{r^2} + g \delta_{i j}\right] 
\eea
$f$ and $g$ are, respectively, longitudinal and lateral correlation functions, which are
\bea
\ds f(r)= \frac{\left\langle u_r({\bf x}) u_r({\bf x}+{\bf r}) \right\rangle }{u^2}, \
\ds g(r)= \frac{\left\langle u_n({\bf r}) u_n({\bf x}+{\bf r}) \right\rangle }{u^2}
\eea
where $u_r$ and $u_n$ are, respectively, the velocity components parallel and normal to  $\bf r$, whereas $r = \vert {\bf r} \vert$ and 
$u^2$ = $\left\langle u_r^2 \right\rangle$ =$\left\langle u_n^2 \right\rangle$=
$1/3 \left\langle u_i u_i \right\rangle $.
Due to the continuity equation, $f$ and $g$ are linked each other by the relationship
\bea
g = f + \frac{1}{2}  \frac{\partial f}{\partial r} r
\label{g}
\eea

The von K\'arm\'an-Howarth equation reads as follows \cite{Karman38, Batchelor53}
\bea
\ds \frac{\partial f}{\partial t}  = 
\ds \frac{K}{u^2}    +
\ds 2 \nu  \left(  \frac{\partial^2 f} {\partial r^2} +
\ds \frac{4}{r} \frac{\partial f}{\partial r}  \right)  
%- \frac{1}{u^2} \frac{d u^2}{d t} f 
-    10 \nu \frac{\partial^2 f}{\partial r^2}(0) f
\label{vk}  
\eea
where $K$ is an even function of $r$, which is defined by the following equation \cite{Karman38, Batchelor53}
\bea
\left(    r \frac{\partial}{\partial r}  + 3  \right) K(r) =
\ds  \frac{\partial }{\partial r_k} 
\ds  \left\langle u_i  u_i' (u_k - u_k')  \right\rangle 
\label{vk1}  
\eea
and which can also be expressed as
\bea
K(r)= u^3 \left(  \frac{\partial}{\partial r}  + \frac{4}{r}  \right) k(r)
\label{kk}
\eea
where $k$ is the longitudinal triple correlation function
\bea
\ds k(r)= \frac{\left\langle u_r^2({\bf x}) u_r({\bf x}+{\bf r}) \right\rangle }{u^3}
\eea

The boundary conditions of Eq. (\ref{vk}) are \cite{Karman38, Batchelor53}
\bea
f(0) = 1, \ \ \lim_{r \rightarrow \infty} f(r) = 0
\label{bc}
\eea
The viscosity is responsible for the decay of the turbulent kinetic energy, the rate of which is
\cite{Karman38, Batchelor53}
 %obtained putting $r=0$ in the von K\'arm\'an-Howarth equation, i.e.
\bea
\frac{d u^2} {d t} = 10 \nu u^2 \frac{\partial^2 f}{\partial r^2}(0) 
\label{ke}
\eea
This energy is distributed at different wave-lengths according to the energy spectrum 
$E(\kappa)$ which is calculated as the Fourier Transform of $f u^2$, whereas 
the ''transfer function'' $T(\kappa)$ is the Fourier Transform of $K$ \cite{Batchelor53}, i.e.
\bea
\hspace{-1.0mm}
\left[\begin{array}{c}
\hspace{-1.0mm} \ds E(\kappa) \\\\
\hspace{-1.0mm} \ds T(\kappa)
\end{array}\right]  
\hspace{-1.5mm}= 
\hspace{-1.5mm} \frac{1}{\pi} 
\hspace{-1.0mm} \int_0^{\infty} 
\hspace{-1.5mm}\left[\begin{array}{c}
\hspace{-1.0mm} \ds  u^2 f(r) \\\\
\hspace{-1.0mm} \ds K(r)
\end{array}\right]  \kappa^2 r^2 \hspace{-1.0mm} 
\left( \hspace{-0.5mm}\frac{\sin \kappa r }{\kappa r} - \cos \kappa r \hspace{-0.5mm} \right) d r 
\label{Ek}
\eea
where $\kappa = \vert {\bf \bfkappa} \vert$ and $T(\kappa)$ identically 
satisfies to the integral condition
\bea
\int_0^\infty T(\kappa) d \kappa = 0
\label{tk0}
\eea 
which states that $K$  does not modify the total kinetic energy.
The rate of energy dissipation $\varepsilon$ is calculated for 
isotropic turbulence as follows \cite{Batchelor53}
\bea
\ds \varepsilon = -\frac{3}{2} \frac{d u^2} {d t}=
  2 \nu \int_0^{\infty} \kappa^2 E(\kappa) d \kappa
\eea
The microscales of Taylor $\lambda_T$, and of Kolmogorov $\ell$, are defined as
\bea
\begin{array}{c@{\hspace{+0.2cm}}l}
\ds \lambda_T^2 = \frac{u^2}{\langle (\partial u_r/\partial r)^2 \rangle} = 
-\frac{1}{\partial^2 f/\partial r^2(0)}, \ 
\ds \ell = \left( \frac{\nu^3} { \varepsilon}\right)^{1/4}
\end{array}
\eea

\bigskip

\section{\bf Appendix B: Critical Reynolds number \label{s3} }

The purpose of this appendix is to provide an estimation of the critical Reynolds number
assuming that the turbulence is, in any case, fully developed,
homogeneous and isotropic. 
Thus, the obtained results are subjected to these assumptions.

To this end, consider now the equation of motion of a fluid particle
$
{d {\bf x}} / {dt} = {\bf u} ({\bf x}, t)
$
and its fixed points which satisfy  
${d {\bf x}} / {dt} = 0$.
We assume that the bifurcations cascade of this equation are expressed in terms of the
characteristic scales by the asymptotic approximation \cite{Guckenheimer90}
\bea
l_n = \frac{l_{1}} {\alpha^{n-1}}
\label{scales0}
\eea
where $\alpha$ $\approx 2$ \cite{Guckenheimer90, Feigenbaum78}, and $l_n$ represent the average distance between two branches of fixed points which born in the same bifurcation.
Equation (\ref{scales0}) is supposed to describe the route toward the chaos and is assumed to be valid until the onset of the turbulence. 
In this situation the minimum for $l_n$ can not be less than the dissipation length or Kolmogorov scale $\ds \ell = (\nu^3/ \varepsilon)^{1/4}$ \cite{Landau44}, where $l_1$ gives a good estimation of the correlation length of the phenomenon \cite{Guckenheimer90, Prigogine94}
which, in this case is the Taylor scale $\lambda_T$.
Thus, $\ell < l_n < \lambda_T$, and
\bea
\ds \ell = \frac{\lambda_T}  {\alpha^{N-1}}
\label{scales01}
\eea
where $N$ is the number of bifurcations at the beginning of the turbulence.
Equation (\ref{scales01}) gives the connection between 
the critical Reynolds number and number of bifurcations.
In fact, the characteristic Reynolds numbers associated to the scales 
$\ell$ and $\lambda_T$ are $R_K = \ell u_K/\nu \equiv$ 1 and
 $R_{\lambda} = \lambda_T u/\nu$, respectively, where 
$\ds u_K = (\nu \varepsilon )^{1/4}$ is characteristic velocity at the Kolmogorov scale,
 and $u =\sqrt{\left\langle u_i u_i \right\rangle/3}$ is the velocity standard
deviation \cite{Batchelor53}.
For isotropic turbulence, these scales are linked each other by \cite{Batchelor53}
\bea
\ds {\lambda_T}/{\ell} = 15^{1/4} \sqrt{R_\lambda}
\label{scales}
\eea
In view of Eq. (\ref{scales01}), this ratio can be also expressed through $N$, i.e.
\bea
\alpha^{N-1} = 15^{1/4} \sqrt{R_\lambda}
\label{scales1}
\eea
Assuming that $\alpha$ is equal to the Feigenbaum constant ($2.502...$),
the value $R_\lambda \simeq$ 1.6 obtained for $N=$ 2 is not compatible with $\lambda_T$
which is the correlation scale, while the result $R_\lambda \simeq$ 10.12, calculated for
$N=$ 3, is an acceptable minimum value for $R_\lambda$. The order of magnitude of these values 
can be considered in agreement with the various scenarios describing the roads to the turbulence \cite{Ruelle71, Feigenbaum78, Pomeau80, Eckmann81}, and with the diverse 
experiments \cite{Gollub75, Giglio81, Maurer79} which state that the turbulence begins 
for $N \ge 3$.
Of course, this minimum value for $R_\lambda$ is the result of the assumptions 
$\alpha \simeq$ 2.502, $l_1 \simeq \lambda_T$, $l_N \simeq \ell$ and of the asymptotic approximation (\ref{scales0}).


\end{document}